\documentclass{AIAA}
\usepackage{amssymb}
\usepackage{color}
\usepackage{longtable}

\begin{document}

\title{Wavelet-based Adaptive Techniques Applied to Turbulent Hypersonic Scramjet Intake Flows}

\author{Sarah Frauholz\footnote{Research Fellow, Chair for Computational Analysis of Technical Systems, Center for Computational Engineering Science (CCES), RWTH Aachen, Schinkelstr. 2, 52062 Aachen, frauholz@cats.rwth-aachen.de.}} \author{Arianna Bosco\footnote{Aerodynamics Developer Engineer, MTU Aero Engines, Munich.}}
\author{Birgit U.~Reinartz\footnote{Research Associate, Chair for Computational Analysis of Technical Systems (CATS), CCES, Senior Member AIAA, RWTH Aachen, Schinkelstr. 2, 52062 Aachen, reinartz@cats.rwth-aachen.de.}}
\author{Siegfried M\"uller\footnote{Professor, Institut f\"ur Geometrie und Praktische Mathematik, RWTH Aachen, Templergraben 55, 52062 Aachen.}}
\author{Marek Behr\footnote{Professor, Chair for Computational Analysis of Technical Systems (CATS), CCES, RWTH Aachen, Schinkelstr. 2, 52062 Aachen, behr@cats.rwth-aachen.de.}}
\affiliation{RWTH Aachen University, Germany} 

\begin{abstract}
The simulation of hypersonic flows is computationally demanding due to the large gradients of the flow variables at hand, caused both by strong shock waves and thick boundary or shear layers. The resolution of those gradients imposes the use of extremely small cells in the respective regions. Taking turbulence into account intensifies the variation in scales even more. 
 Furthermore, hypersonic flows have been shown to be extremely grid sensitive. For the simulation of fully three-dimensional configurations of engineering applications, this results in a huge amount of cells and as a consequence prohibitive computational time. Therefore, modern adaptive techniques can provide a gain with respect to both computational costs and accuracy, allowing the generation of locally highly resolved flow regions where they are needed and retaining an otherwise smooth distribution. \\
In this paper, an $h$-adaptive technique based on wavelets is employed for the solution of hypersonic flows. The compressible Reynolds averaged Navier-Stokes equations are solved using a differential Reynolds stress turbulence model, well suited to predict shock-wave-boundary-layer interactions in high enthalpy flows. Two test cases are considered: a compression corner at 15 degrees and a scramjet intake. The compression corner is a classical test case in hypersonic flow investigations because it poses a shock-wave-turbulent-boundary-layer interaction problem. 
The adaptive procedure is applied to a two-dimensional configuration as validation. The scramjet intake is firstly computed in two dimensions. Subsequently a three-dimensional geometry is considered. Both test cases are validated with experimental data and compared to non-adaptive computations. The results show that the use of an adaptive technique for hypersonic turbulent flows at high enthalpy conditions can strongly improve the performance in terms of memory and CPU time while at the same time maintaining the required accuracy of the results.
\end{abstract}

\maketitle

\section*{Nomenclature}
\noindent\begin{longtable}{@{}lcl@{}}
$c_{p}$ & : & Specific heat at constant pressure, pressure coefficient [-]\\
$\delta_{ij}$ & : & Kronecker Delta [-]\\
$E$ & : & Specific total energy [m$^2$/s$^2$]\\
$\epsilon$ & : & Turbulence dissipation rate [m$^2$/s$^3$] \\
$\varepsilon$& : & Threshold value used for data compression [-]\\
$\varepsilon_l$& : & Level-dependent threshold value for level l [-]\\
$H$ & : & Total specific enthalpy [m$^2$/s$^2$]\\
$k$ & : & Turbulent kinetic energy [m$^2$/s$^2$]\\
$II$& : & Second invariant of the anisotropy tensor [-]\\
$L$ & : & Maximal refinement level [-]\\
$l$ & : & Local refinement level [-]\\
$\mu$ & : & Molecular viscosity [kg/(m s)]\\
$\omega$ & : & Specific turbulence dissipation rate [1/s]\\
$p$ & : & Pressure [Pa]\\
$p_t$ & : & Total pressure [Pa]\\
$q_{i}$ & : & Component of heat flux vector [W/m$^2$] \\
$q_k^{(t)}$ & : & Turbulent heat flux [W/m$^2$] \\
$\rho$ & : & Density [kg/m$^3$]\\
$St$& : & Stanton number [-]\\
$t$ & : & Time [s]\\
$T$& : & Temperature [K]\\
$T_{\rm w}$ & : & Wall temperature [K] \\
$T_{0}$ & : & Total temperature [K] \\
$U$ & : & Local velocity [m/s] \\
$u_i$& : & Velocity component [m/s]\\
$x_i$& : & Cartesian coordinates component [m]\\
$x$, $y$, $z$ & : & Cartesian coordinates [m]\\
$y^+$ & : & Dimensionless wall distance [-]\\
$M$ & : & Mach number [-]\\
$Re$ & : & Reynolds number [1/m]\\
$b_{ij}$ & : & Anisotropy tensor [-]\\
$D_{ij}$ & : & Diffusion tensor for the Reynolds stresses [m$^2$/s$^3$]\\
$\epsilon_{ij}$& : & Destruction tensor for the Reynolds stresses [m$^2$/s$^3$]\\
$M_{ij}$ & : & Turbulent mass flux tensor for the Reynolds stresses [m$^2$/s$^3$]\\
$\Pi_{ij}$ & :  & Re-distribution tensor for the Reynolds stresses [m$^2$/s$^3$]\\
$P_{ij}$ & : & Production tensor for the Reynolds stresses [m$^2$/s$^3$]\\
$S_{ij}$ & : & Strain rate tensor [1/s]\\
$\tilde{R}_{ij}$ & : & Reynolds stress tensor [m$^2$/s$^2$]\\
$\tau_{ij}$ & : & Viscous stress tensor [m$^2$/s$^2$]\\
$W_{ij}$ & : & Rotation tensor [1/s] \\
$\frac{\partial \cdot}{\partial \cdot}$ & : & Partial derivative \\
$\overline{\cdot}$ & : & Reynolds-averaged quantity \\
$\tilde{\cdot}$ & : & Favre-averaged quantity \\
$\cdot_\infty$ & : & Free stream value 
\end{longtable}

\section{Introduction}
\label{intro}

The study of hypersonic flows has been of interest for more than 50 years \cite{martin}. Nowadays, a major application in the field of hypersonics is the realization of a supersonic combustion ramjet (scramjet), an airbreathing propulsion system that operates above Mach 5 and at approximately 30--40 km altitude. One major impediment to the realization of such an engine lies in the uncertainties related to its aerothermodynamic design. The study of hypersonic configurations at real flight conditions is both experimentally as well as numerically demanding, though not for the same reasons. On the one hand, hypersonic test facilities need a huge amount of energy to establish high-enthalpy flow conditions. Short duration test times and vitiated air effects are just two of the resulting drawbacks. On the other hand, numerical simulations have to deal with modeling uncertainties with respect to turbulence and high temperature effects as well as limited computer resources. As the models grow more sophisticated, computer power is stretched to its limits. In hypersonic viscous flow, this situation is paired with an additional, unusually demanding requirement on grid resolution \cite{Cand:00, Rein:07b}. 

In order to obtain high accuracy but at an affordable computational cost, adaptive strategies can be employed. These techniques aim at the reduction of the size of the discrete problem by locally refining the mesh in action regions of the flow while keeping it coarse elsewhere.
Adaptive techniques applied to compressible turbulent flows have been successfully used for aerothermodynamic applications in the past, e.g., in the DLR TAU code \cite{Mack:02, tau2}. In \cite{tau2}, it is stressed that the performance of the adaptive procedure is \emph{strongly dependent on the initial grid} which has to be of sufficient resolution and quality. When such an initial resolution is not used, a clustering of cells in the wrong flow regions occurs. Thus, to successfully apply the adaptation procedure, a priori knowledge of the flow is required. This knowledge becomes exceedingly hard to come by for 
complex flows.

In this context, the mathematical concept of multiresolution-based grid adaptation plays a central role in that it self-reliably detects all physical 
relevant effects and resolves them reasonably even when starting from a \emph{coarse} grid.
First work in this regard has been published in \cite{GottschlichMueller-Mueller:99} motivated by Harten's work \cite{harten}. The basic idea is to perform a multiscale analysis of a sequence of cell averages associated with any finite volume discretization on a given highest level of resolution (reference mesh). This results in cell averages on some coarse level and the fine scale information is encoded in arrays of detail coefficients of ascending resolution. Subsequently, threshold techniques are applied to the multiresolution decomposition where detail coefficients below a threshold value are discarded. By means of the remaining significant details, a locally refined mesh is obtained and its complexity is substantially reduced in comparison to the underlying reference mesh. For a detailed review on multiresolution-based grid adaptation we refer to the monograph by \cite{mueller:09} and the references cited there.\\

The main aim of this paper is to show the possibility of using such an adaptive technique for hypersonic turbulent flow for fully three-dimensional configurations, using as a starting point a truly coarse grid and relying on the adaptive procedure to identify the flow regions where grid refinement is necessary. In this work, the adaptive and parallel solver QUADFLOW \cite{Bramkamp:04} is used. This solver has been designed as an integrated tool in a way that each of its constituents, namely the flow solver, the grid generation, and the grid adaptation, support each other to the highest possible extent.
Specifically, the core ingredients are:
(i) the flow solver concept based on a finite volume discretization \cite{Bramkamp:03b}, 
(ii) the grid generator based on B-spline mappings defined on a multi-block topology \cite{SM-Lamby:07}, and
(iii) the grid adaptation concept based on wavelet techniques \cite{Mueller:03}. 
These three constituents do not solely work together as black boxes which communicate only via interfaces. On the contrary, they have been designed as one program package to efficiently solve aerodynamic problems with a wide variation of scales. Recently, this solver has been parallelized using the concept of space-filling curves \cite{Brix:09,Brix:11}.

To this day, turbulent flow simulations for engineering applications at realistic flight Reynolds numbers are only computationally affordable when applying the Reynolds Averaged Navier-Stokes (RANS) equations. The most widely used turbulence models in this field are the eddy viscosity models, where a linear dependence between the Reynolds stress tensor and the strain rate tensor is assumed. However, several literature reviews showed that these models perform poorly for wall dominated flows characterized by thick boundary layer, strong shock-wave-boundary-layer interaction and separation \cite{roy-blottner}, as are typical for hypersonic applications. For this reason, a differential Reynolds stress turbulence model (RSM) has been preferred in this work. This class of models has not been widely used because of its decreased stability and the increased computational cost due to the presence of seven equations that describe turbulence. However, in an earlier study, the RSM was successfully used for the simulation of separated hypersonic boundary layer flow where common two-equations eddy viscosity models failed \cite{Bosco:11b, Bosco:11}. So far, complex three-dimensional computations with engineering applications have only been performed using the differential Reynolds stress model on block-structured, non-adaptive grids; in the current study, we will show its application to adaptive grids as well.  

In the following, the physical modeling with a special emphasis on the chosen turbulence model is shortly described in Section \ref{pm}. Subsequently, the numerical methods employed for the solution of the discrete problems are illustrated in Section \ref{qf}. Finally in Section \ref{results}, the numerical results are presented. The computations are validated with experimental data. 
The two-dimensional computations are used to evaluate the parameters steering the adaptation process and to assess the possible speed-up of an adaptive simulation vs. a simulation based on a uniform grid.
 Then the established procedure is applied to a fully three-dimensional test case and it is shown that the performance improvement as well as the computational accuracy of the results are maintained. 

\section{Physical Modeling}\label{pm}

In this work, the compressible Reynolds Averaged Navier-Stokes (RANS) equations are solved which describe the conservation of mass, momentum and energy for compressible turbulent flows. The RANS equations read as follows:

\begin{equation}
\frac{\partial \overline{\rho}}{\partial t} + \frac{\partial}{\partial x_k}( \overline{\rho} \tilde{u}_k )=0 \;\;,
\end{equation}

\begin{equation}\label{mom_av}
\frac{\partial}{\partial t}(\overline{\rho}\tilde{u}_i) + \frac{\partial}{\partial x_k}(\overline{\rho}\tilde u_i \tilde u_k) + \frac{\partial}{\partial x_k}(\overline{\rho}\tilde R_{ik})= -\frac{\partial \overline{p}}{\partial x_i} + \frac{\partial \overline{\tau}_{ik}}{{\partial x_k}}\;\;,
\end{equation}

\begin{equation}\label{energy_av}
\frac{\partial}{\partial t} (\overline{\rho}\tilde E) + \frac{\partial}{\partial x_k}(\overline{\rho}\tilde H \tilde u_k)+\frac{\partial}{\partial x_k}(\overline{\rho}\tilde{R}_{ik}\tilde u_i) = \frac{\partial}{\partial x_k}(\overline {\tau}_{ik}\tilde u_i) - \frac{\partial \overline q_k}{\partial x_k} + \overline{\rho} D^{(k)} - \frac{\partial q^{(t)}_k}{\partial x_k}\;\;.
\end{equation}\\

The standard notation for the Reynolds average ($\bar{\cdot}$) and Favre average ($\tilde{\cdot} $) is employed.
The system of equations is closed using the perfect gas assumption, the Fourier assumption for the laminar and turbulent heat fluxes and the assumption of Newtonian fluid for the laminar viscous stresses. The turbulent closure is described below.

\subsection{SSG/LRR-$\omega$ Turbulence Model}\label{ssg-lrr}

For the simulations presented in this work, a differential Reynolds stress model has been chosen as closure for the RANS equations.
The SSG/LRR-$\omega$ model by Eisfeld \cite{Eisfeld:06} is a combination of two previously existing models: The Speziale, Sarkar and Gatski (SSG) model \cite{SSG} using an $\epsilon$-based length scale equation is employed in the far field and coupled to the $\omega$-based Launder, Reece and Rodi (LRR) model \cite{LRR} in its modified Wilcox version \cite{wi_1} for the near wall region.  This was done in order to employ each model in the region where it performs best. 
On the one hand, an $\epsilon$-based model is preferred away from
the wall to avoid the high sensitivity to free-stream turbulence observed in $\omega$-based models.
On the other hand, the choice of an $\omega$-based model near the wall is justified by the desire of
having a low Reynolds number model allowing integration up to the wall. As Wilcox \cite{wi_1} shows,
the near-wall behavior of second-order closure models is strongly influenced by the scale-determining equation. Models based on an $\omega$-equation often predict acceptable values
of the wall integration constant and are quite easy to integrate through the viscous sublayer
compared to models based on an $\epsilon$-equation. Here, the $\omega$-equation by Menter \cite{me_1} is employed to provide the turbulent length scale. Consequently, the blending between the two models is performed using the Menter blending function (\ref{blending})
as well. 

The Reynolds stress tensor is defined as
\begin{equation}
	\bar{\rho}\tilde{R}_{ij} = \overline{\rho u''_iu''_j} \quad.
\end{equation} 
The transport equations for the Reynolds stresses read as follows:
\begin{equation}
	\frac {\partial}{\partial t}(\bar{\rho}\tilde{R}_{ij} )+\frac{\partial}{\partial{x_k}}(\bar{\rho}\tilde{U}_k\tilde{R}_{ij}) = 
\bar{\rho}P_{ij}+\bar{\rho}\Pi_{ij}-\bar{\rho}\epsilon_{ij}+\bar{\rho}D_{ij}+\bar{\rho}M_{ij} \quad .
\end{equation} 

The terms on the right hand side of the equation represent the production, re-distribution, destruction, diffusion and the contribution of the turbulent mass flux, respectively. Apart from the production term, which is exact, all other terms need to be modeled.

The production term defines the interchange of kinetic energy between the mean flow and the fluctuations:
\begin{equation}
\bar{\rho} P_{ij}= -\bar{\rho}\tilde{R}_{ik} \frac{\partial\tilde{u}_j}{\partial x_k}
-\bar{\rho}\tilde{R}_{jk} \frac{\partial\tilde{u}_i}{\partial x_k}.
\end{equation}

The re-distribution term is modeled as follows:
\begin{equation}\label{red}
\bar{\rho}\Pi_{ij}= -(C_1\bar{\rho}\epsilon+\frac{1}{2}C_1^*\bar{\rho}P_{kk})\tilde b_{ij} 
+C_2\bar{\rho}\epsilon(\tilde b_{ik}\tilde b_{kj}-\frac{1}{3}\tilde b_{mn}\tilde b_{mn}\delta_{ij})
+(C_3-C_3^*\sqrt{II})\bar{\rho}\tilde k \tilde S^*_{ij}
\end{equation}
$$
+C_4 \bar{\rho}\tilde k (\tilde b_{ik}\tilde S_{jk}+\tilde b_{jk}\tilde S_{ik}- \frac{2}{3}\tilde b_{mn}\tilde S_{mn}\delta_{ij}) +C_5\bar{\rho}\tilde k(\tilde b_{ik} \tilde W_{jk} + \tilde b_{jk} \tilde W_{ik}) \quad ,
$$
where all the coefficients are obtained inserting the values in Table \ref{coef} in the blending function:
\begin{equation}\label{blending-function}
\phi = F \phi^{LRR} + (1 - F)\phi^{SSG} .
\end{equation}

\noindent The blending function of Menter is defined as:
\begin{equation}\label{blending}
F = \rm tanh (\zeta^4)\;,\;\;\;\;  \zeta = \rm min \left(  \rm max \left( \frac {\sqrt{\tilde k}}{C_{\mu}\omega d}; \frac{500 \bar{\mu}}{\bar{\rho}\omega d^2}\right) ; \frac{4 \sigma_{\omega}^{(SSG)}\bar{\rho}\tilde{k}}{\bar{\rho}C_D^{(SSG)}d^2}  \right).
\end{equation}

\begin{table}[!ht] 
\begin{center} 
\caption{Coefficients of SSG and LRR model for the re-distribution term.}\label{coef}
\medskip 
\begin{tabular}{ c c c c c c c c }
\noalign{\vspace{4pt}}
    & $C_1$ & $C_1^*$ & $C_2$  & $C_3$ & $ C_3^*$ & $C_4$ & $C_5$\\ \hline
    SSG & 3.4 & 1.8 & 4.2 & 0.8 &1.3 & 1.25 & 0.4\\ \hline
    LRR & 3.6 & 0 & 0 & 0.8 & 0 & 2.0 & 1.11\\
\end{tabular} 
\end{center} 
\end{table}

In the above equation, $\tilde k$ is the (specific) turbulent kinetic energy and $\epsilon$ is the isentropic dissipation rate defined as follows:
\begin{equation}
\tilde k = \frac{\tilde R_{kk}}{2} \;\;\;,\;\; \epsilon=C_{\mu}\tilde k \omega ,
\end{equation}
\noindent where $C_{\mu}$ = 0.09.

The tensors appearing in equation (\ref{red}) are the anisotropy tensor 
\begin{equation}
\tilde b_{ij} = \frac{\tilde R_{ij}}{2\tilde k} - \frac{\delta_{ij}}{3} ,
\end{equation}

\noindent and $II=\tilde{b}_{ij}\tilde{b}_{ij}$ its second invariant, the strain rate tensor and the rotation tensor

\begin{equation}
\tilde S_{ij} = \frac{1}{2} \left( \frac{\partial \tilde U_i}{\partial x_j}+ \frac{\partial \tilde U_j}{\partial x_i} \right) \;\;,\;\; \tilde W_{ij} = \frac{1}{2} \left( \frac{\partial \tilde U_i}{\partial x_j}- \frac{\partial \tilde U_j}{\partial x_i} \right) ,
\end{equation}

\noindent and the traceless strain rate tensor
\begin{equation}
\tilde S^*_{ij} = \frac{1}{2} \left( \frac{\partial \tilde U_i}{\partial x_j}+ \frac{\partial \tilde U_j}{\partial x_i} \right) - \frac{1}{3} \frac{\partial \tilde U_k}{\partial x_k} \delta_{ij}  .
\end{equation}
\noindent The isotropic destruction term is:
\begin{equation}
\bar{\rho}\epsilon_{ij} = \frac{2}{3}C_{\mu}\bar{\rho}\tilde k \omega \delta_{ij}  .
\end{equation}

\noindent For the diffusion term, the generalized gradient diffusion model is chosen:
\begin{equation}
\bar{\rho}D_{ij} = \frac{\partial}{\partial x_k}\left(\left( \bar{\mu}\delta_{kl}+ D^{(GGD)}\frac{\rho}{\omega}\tilde{R}_{kl}\right)\frac{\partial \tilde R_{ij}}{\partial x_l} \right) .
\end{equation}

\noindent The value of the constant $D^{(GGD)}$ is computed by the equation:
\begin{equation}
D^{(GGD)} = F\sigma^* + (1-F)\frac{C_s}{C_{\mu}} .
\end{equation}

\noindent $F$ is the blending equation in (\ref{blending}), $\sigma^*$= 0.5 and $C_s$=0.22.

\noindent Finally the term $\bar{\rho}M_{ij}$ is neglected because no good model for it exists yet.

\section{Numerical Methods}\label{qf}

\subsection{QUADFLOW Solver}
QUADFLOW is a well validated flow solver which solves the RANS equations for unsteady, compressible fluid flow in two and three dimensions \cite{Bramkamp:04}. This solver has been developed over a period of more than one decade within the Collaborative Research Center SFB 401 {\em Modulation of Flow and Fluid-Structure Interaction at Airplane Wings}~\cite{Ballmann:03,schroeder} at RWTH Aachen University. The flow solver in QUADFLOW is based on a cell-centered finite volume discretization. The mesh is treated as fully unstructured and composed of polygonal (2D) or polyhedral (3D) elements. This approach is especially suited for dealing with hanging nodes appearing in locally adaptive meshes. For the time and space discretization the user can choose among several options for the Riemann solver, the limiter, the reconstruction and the Runge-Kutta scheme. Here we summarize the methods used for the computations presented in Section \ref{results}:
The convective fluxes are discretized using the AUSMDV Riemann solver. A linear reconstruction of the primitive variables is performed to locally achieve second order accuracy in space, and the Venkatakrishnan slope limiter is employed to avoid oscillations typical of higher order schemes \cite{limiter}. For the discretization of the viscous fluxes, a modified central difference method is used \cite{Bramkamp:03}. 
A second order accurate explicit Runge-Kutta scheme is employed for the time integration. For the treatment of turbulent flows, the user can choose from a wide variety of eddy viscosity models, one explicit algebraic Reynolds stress model and a differential Reynolds stress model that is used in the current study.

At the far field boundaries, supersonic inflow or outflow conditions have been imposed. At solid boundaries the no-slip condition and an isothermal wall have been prescribed. Concerning the turbulent variables, the no-slip condition also implies that the Reynolds stresses are zero at the wall. The chosen $\omega$-wall condition is the one from Menter \cite{me_1} imposing a value of this quantity depending on the distance of the first cell center from the wall. For three-dimensional simulations of the intake a half model is used and a symmetry condition is imposed at one side.

\subsection{Adaptive Technique}\label{adaptive}
\begin{figure*}[h!]
  \centerline{\includegraphics [width = 1.0\textwidth,clip]{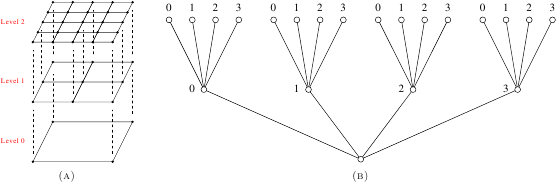}}
        \caption{Hierarchy of nested Cartesian grids (A) as well as corresponding tree (B).}\label{grid-adap-1}
  \end{figure*}

The main distinction from previous works lies in the fact that here recent multiresolution techniques based on biorthogonal wavelets are employed~\cite{harten96,carnicer}. The starting point is to transform the arrays of cell averages associated with any given finite volume discretization into a different format that reveals insight into the characteristic contributions of the solution to different length scales. The cell averages on a given highest level of resolution $l=L$ are represented as cell averages on some coarse level $l=0$, while the intermediate fine scale information is encoded in arrays of detail coefficients of ascending resolution $l=0,\ldots,L-1$. This requires a hierarchy of meshes as exemplified by
 Figure \ref{grid-adap-1}.

\begin{figure*}[h!]
 \centerline{\includegraphics [width = 1.0\textwidth,clip]{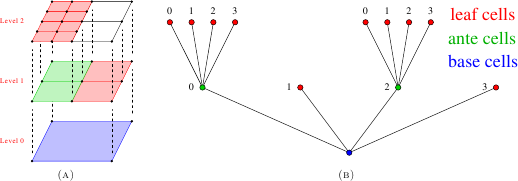}}
        \caption{Grid adaptation procedure in case of Cartesian grids (A) as well as corresponding graded tree (B).}\label{grid-adap-2}
  \end{figure*}

 The multiscale representation is used to create locally refined meshes proceeding in the following three steps \cite{Mueller:03}:
\newline
Step 1. Due to the cancellation property of biorthogonal wavelets the details may become small, if the underlying data are locally smooth. Therefore, quite in the spirit of image compression, the vector of details may be compressed by means of hard thresholding, i.e., all detail coefficients whose absolute values fall below a level-dependent threshold $\varepsilon_l = 2^{(l-L)\bar{d}} \varepsilon$ for a suitable parameter $\varepsilon$, where $\bar{d}$ denotes the spatial dimension, are discarded. Note that the compression rate becomes higher with the number of vanishing moments of the wavelet functions.  Ideally, the threshold value $\varepsilon$ should be chosen such that the {\it perturbation error}, i.e., the difference between the reference solution obtained by the finite volume method (FVM) performed on the uniformly refined grid on level $L$ (reference grid) and the adaptive solution projected onto the reference grid on level $L$, is proportional to the {\it discretization error} of the FVM on the reference mesh. For scalar nonlinear conservation laws rigorous estimates are available \cite{Mueller:2003a, Mueller:2010}.
\newline
Step 2. In order to account for the dynamics of a flow field due to the time evolution, and to appropriately resolve all physical effects on the new time level, this set is to be inflated such that the resulting prediction set contains all $\varepsilon$-significant details of the old \emph{and} the new time level.
The prediction strategy depends on the underlying system of evolution equations to be approximated. Here, Harten's heuristic prediction strategy~\cite{harten} is used.
\newline
Step 3. From the significant details, the locally refined grid and corresponding cell averages are constructed. For this purpose, the grid is considered levelwise from coarse to fine, checking for all cells of a level whether there exists a significant detail. If there is one, the respective cell is refined, i.e., the average of this cell is replaced by the averages of its children by locally applying the inverse multiscale transformation; see Figure \ref{grid-adap-2}.

For the multiresolution analysis, we use biorthogonal wavelets~\cite{Cohen92}. Starting from orthogonal box wavelets, i.e., Haar wavelets~\cite{haar:10} on an arbitrary grid hierarchy, modified biorthogonal wavelets with higher vanishing moments are constructed applying the change of stable completion~\cite{carnicer}. For the computations in Section \ref{results}, we use vanishing moments of order two. A detailed derivation and analysis of the multiscale-based grid adaptation concept and the construction of appropriate biorthogonal wavelets can be found in \cite{Mueller:03}. An overview on recent developments and an extended list of related work is given in \cite{mueller:09}.

Here, we summarize the parameters required by the adaptive procedure used for the computations in Section \ref{results}.   All conservative variables are used to drive the adaptation
process.  Generally, the simulations start on a very coarse grid (e.g., 1000 cells in 2D). An adaptation is performed each time the normalized averaged density residual drops below the drop residual $\varepsilon_{drop}$ (e.g., $10^{-4}$) until the maximum number of adaptations is reached. The normalized averaged density residual is the parameter to measure the steady state convergence. 
The maximum refinement level is chosen such that an additional refinement level does not improve the solution further. Thus, the maximum refinement level is set to $L=5$ for the compression corner and $L=4$ for the scramjet. The threshold value of the respective simulation should be chosen as large as possible to minimize the computational time, while at the same time it has to be small enough to maintain the accuracy by ensuring that the solution of the adapted grid and the uniformly refined grid are superimposed. For the compression corner, the threshold value $\varepsilon=5\times10^{-3}$ is necessary, whereas for the scramjet application the threshold value $\varepsilon=10^{-2}$ is sufficient \cite{Frauholz:12}. 

\section{Results}\label{results}

Within this section, numerical results for a compression corner of 15 degrees in two dimensions are discussed showing the comparison with a uniform grid as well as experimental data and focusing on the performance of the adaptive technique in terms of number of iterations, CPU time and number of cells required. 
Afterwards, a scramjet intake is presented. The simulations are performed in two and three dimensions. The two-dimensional simulation is compared to a non-adaptive structured grid which was specially designed and optimized for this intake configuration during a prior combined numerical and experimental test campaign \cite{Neuen:06, Fischer,Nguyen:10}.  
For the three-dimensional computation, the results are compared to a uniformly refined grid and experimental data in order to show the accuracy and efficiency of the adaptive procedure. 
All two-dimensional computations have been performed on an in-house cluster using 16 processors. The three-dimensional computations were done on the BULL cluster of the RWTH Aachen University with 60 processors.

\subsection{Flow over a compression corner at 15 degrees}
\label{sec:2d-results}%
\begin{figure*}[h!]
 \centerline{\includegraphics [width = 0.9\textwidth,clip]{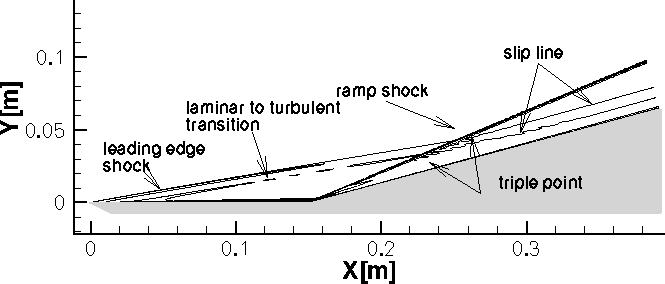}}
        \caption{Mach number contours for a two-dimensional adaptive simulation showing the main physical flow features.} \label{15_deg_mach}
  \end{figure*}

In the field of hypersonics, the flow over a compression corner represents a standard test case, since it combines a straightforward geometry with the two most important physical features: thick turbulent boundary layer and shock-wave-boundary-layer interaction. The flow around a compression corner at 15 degrees is of interest in this section. 
The main physical phenomena occurring along the chosen geometry are depicted in Figure \ref{15_deg_mach}: Along the plate, the leading edge shock wave is visible, as well as the shock generated at the location of the (numerical) transition of the boundary layer from laminar to turbulent. At the kink, a shock wave is generated due to the presence of the compression ramp and it interacts with the two shocks mentioned before giving origin to two slip lines departing from two subsequent triple points.

The inflow conditions used for the numerical simulations are given in Table \ref{inflow}. The flat plate length and the ramp length are the same and are equal to 0.15 m. 

\begin{figure*}[h!]
 \centerline{\includegraphics [width = 0.9\textwidth,clip]{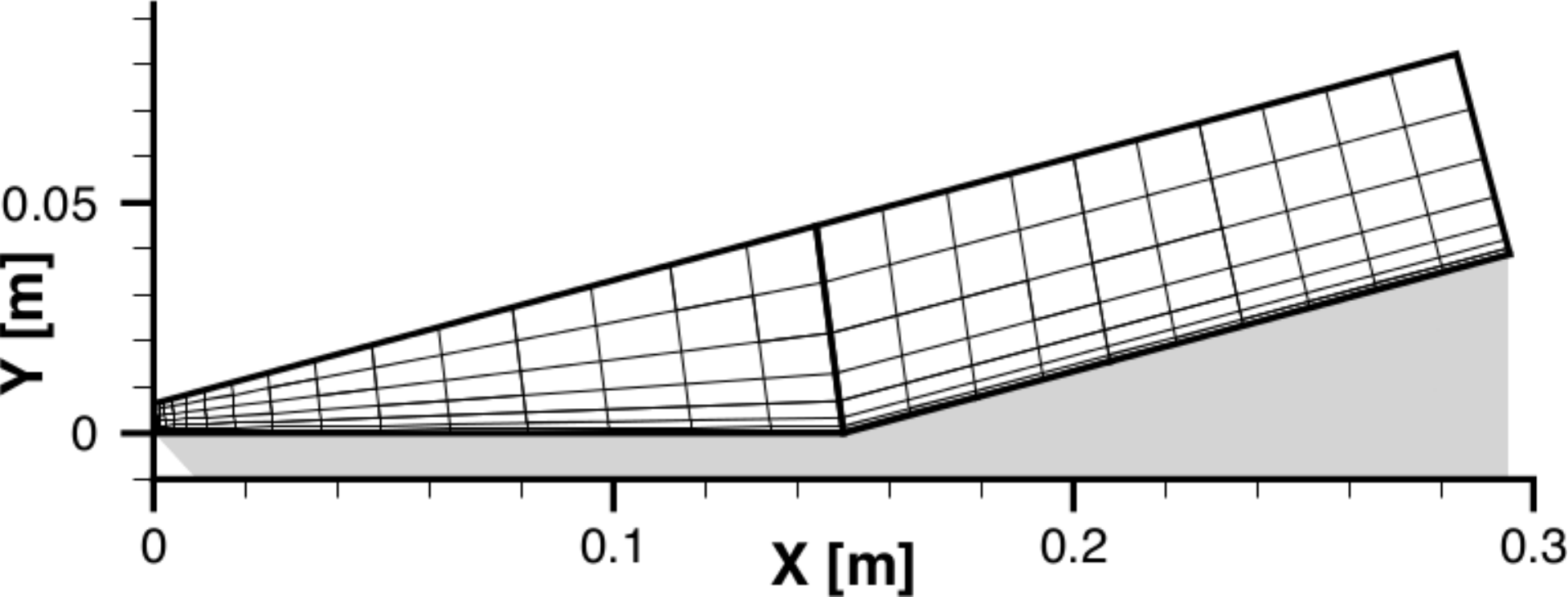}}
 \caption{Initial computational grid at refinement level $L$=1. }\label{grid-L1}
 \end{figure*}

\begin{figure*}[h!]
 \centerline{\includegraphics [width = 0.9\textwidth,clip]{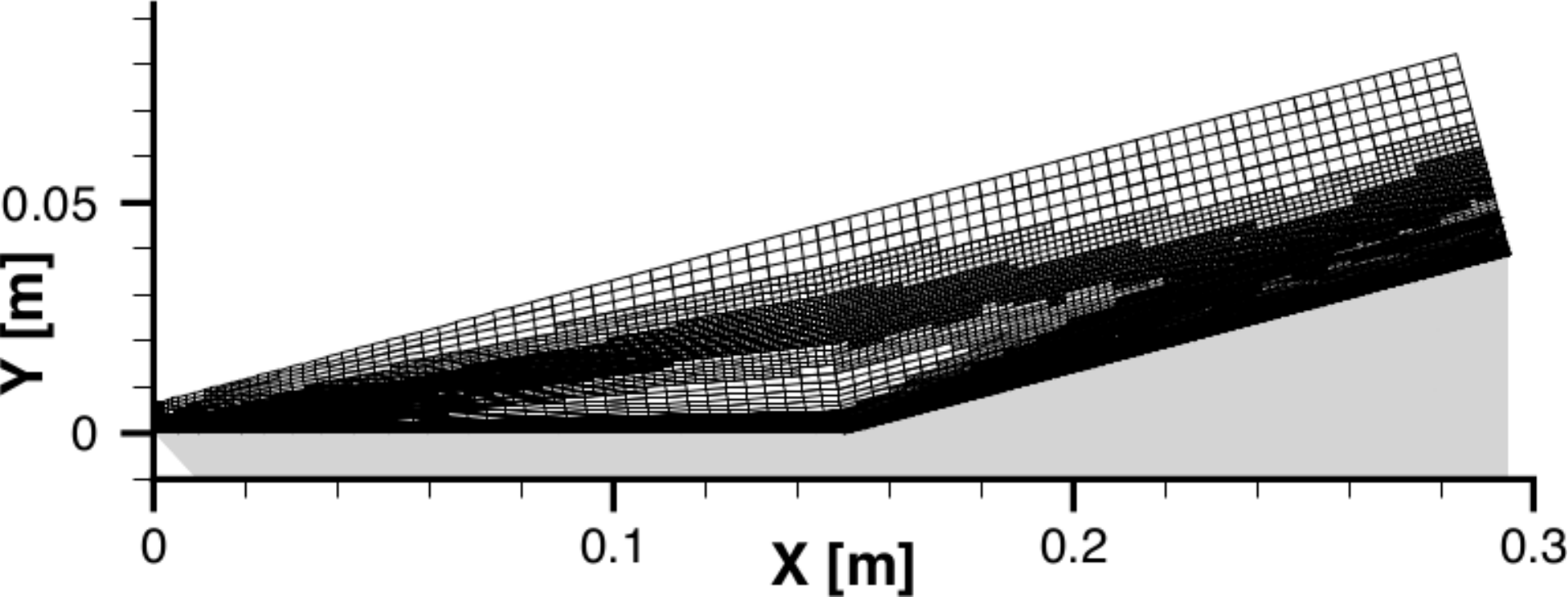}}
\caption{Final computational grid at refinement level $L$=5 after the last adaptation. }\label{grid-L5}
 \end{figure*}

\begin{table*}[h!]
\begin{ruledtabular}
\begin{tabular}{ccccccc}
   $\rho$ [Kg/m$^3$] & $p$ [Pa]  & $U$ [m/s]  &  $M$ [-] & $Re$ [1/m] & $T$ [K] & $T_{\rm w}$ [K]\\ \hline
     0.08624 & 9681 & 2516 &  6.35 &  9.65$\times$10$^6$ &  396 & 300\\
   \end{tabular}
\end{ruledtabular}
 \caption{Test conditions for 15 degrees compression ramp.}\label{inflow}
 \end{table*}
The grid has two blocks and it contains 15 cells in the flow direction and 6 cells in the cross-flow direction at level $L=0$. Cells are clustered near the leading edge and toward the solid wall to obtain the desired resolution of 10$^{-7}$~m on the finest level in these regions. In the other regions the grid is kept as smooth as possible. The initial grid and the final grid after the last refinement are shown in Figure \ref{grid-L1} and Figure \ref{grid-L5}, respectively. Here the refinements triggered by the presence of the boundary layer and the shock waves are clearly visible. Table \ref{adapt} shows the evolution of grid cells during the adaptive procedure.

\begin{table*}[h!]
\begin{ruledtabular}
\begin{tabular}{cccccccc}
  refinement level   & $L$=0 & $L$=1  & $L$=2  &  $L$=3 & $L$=4 & $L$=5 & $L$=5 final\\ \hline
   number of cells   & 90    & 360  & 1,404   & 5,196 & 17,829 &  44,349 & 44,930  \\ 
   \end{tabular}
\end{ruledtabular}
 \caption{Evolution of cell numbers during the adaptive procedure for a two-dimensional compression ramp using five refinement levels.}\label{adapt}
 \end{table*}
%
\begin{figure*}[h!]
\centerline{
\includegraphics [height = 0.5\textwidth,clip]{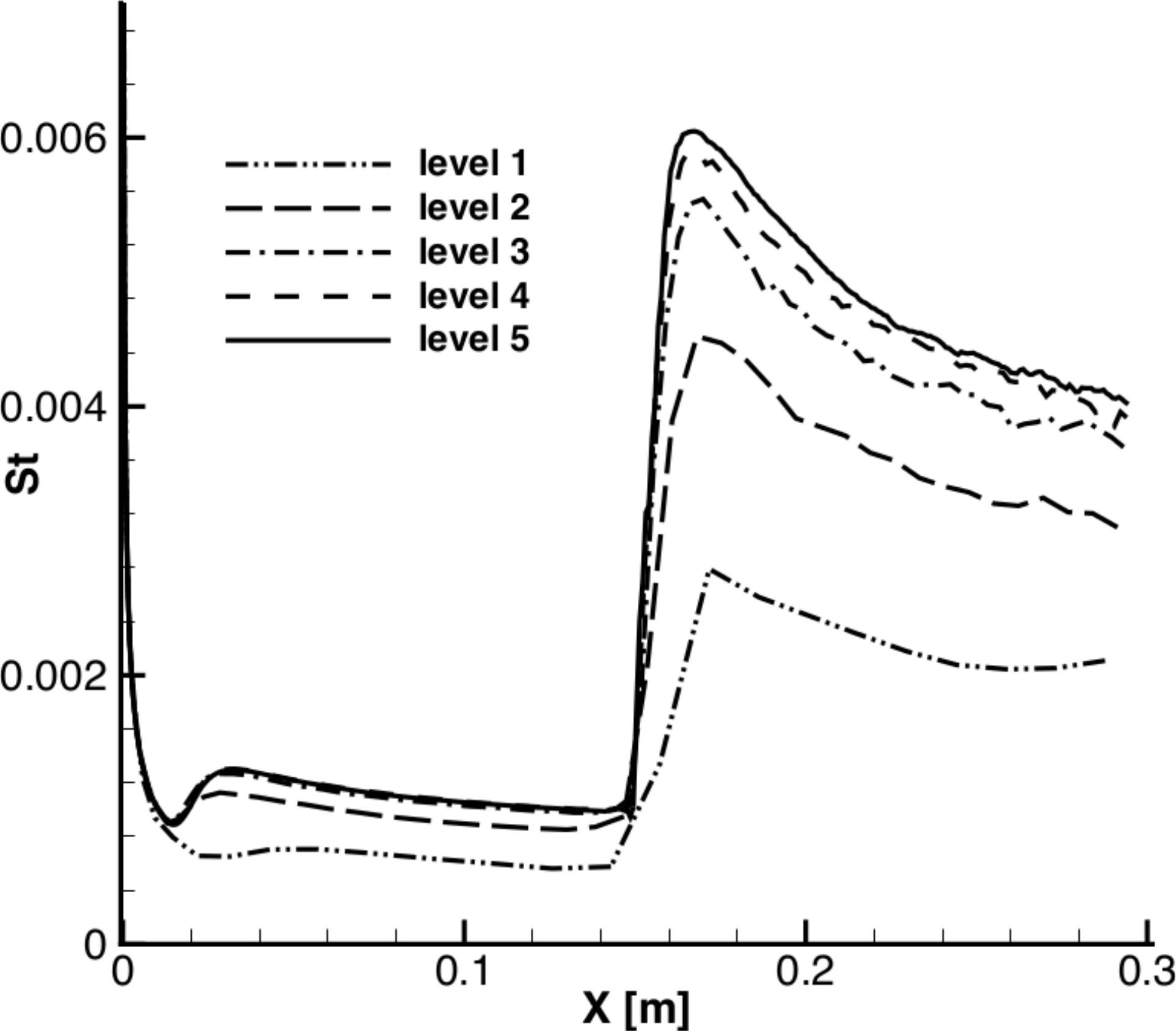}} 
    \caption{Stanton number distribution over a compression corner at 15 degrees. Comparison of different refinement levels during the adaptive computation at averaged density residual of 10$^{-4}$. The highest refinement level is L=5. Inflow conditions: Re= 9.65$\times$10$^{6}$~1/m, M=6.35.} \label{2d-results-a}
\end{figure*}

To illustrate the adaptive procedure, Figure \ref{2d-results-a} presents Stanton number distributions  
\begin{equation}
St = \frac{q_w}{\rho_{\infty} \left| u_{\infty} \right| c_p (T_{0,\infty}-T_w)}
\label{stanton}\end{equation}
\noindent at the wall for different refinement levels of the adaptive computation. 
Since the flow field is initialized by constant data, all details are zero and thus no grid refinement will be triggered. In order to detect physical effects induced by the boundary conditions a uniform refinement of all the grid cells is performed so that the initial grid is always at refinement level $L=1$. 
The level $L=1$ grid is too coarse to resolve the temperature gradient and thus the Stanton number correctly. At the drop residual $\varepsilon_{drop}=10^{-4}$ the adaptation procedure takes place and the adapted level $L=2$ grid is initialized by locally coarsening and refining the solution of the previous level $L=1$ grid. This procedure is repeated until the maximum refinement level $L=5$ is reached. On the final level $L=5$ grid the averaged density residual is dropped until $10^{-6}$. However each adaptation refines the grid and hence the Stanton number resolution improves. As the actual refinement level increases the difference between the solution of the actual and the next refinement level shrinks. Hence with more refinement levels the difference vanishes and the adaptive computation converges against the final solution, where an increase in the number of refinement levels does not improve the solution further. \\

Using adaptive computations we always have to consider two errors: the discretization error and the perturbation error. The discretization error is the difference between the exact, analytic solution and the numerical solution on the fully refined grid (reference grid). The perturbation error is the difference between the numerical solution obtained on the reference grid and the adaptive solution projected onto the reference grid. In general, to show that the discretization error is negligible grid convergences studies are done on a sequence of successively refined grids. However, to show grid convergence of the mesh-adaptive computations in a mathematically correct way it is not sufficient to compare adaptive solutions of two consecutive refinement levels  but the solutions obtained on uniformly refined grids at different refinement levels have to be considered.\\

Therefore, we also computed the solutions of the uniformly refined grids for $L=5$ and $L=6$. Theses two solutions showed no significant difference (not presented here) and thus we consider the uniformly refined $L=5$ solution as grid converged and use it to validate our adaptive results.\\ 


    \begin{figure*}[h!]
  \centerline{
       \mbox{\includegraphics [height = 0.5\textwidth,clip]{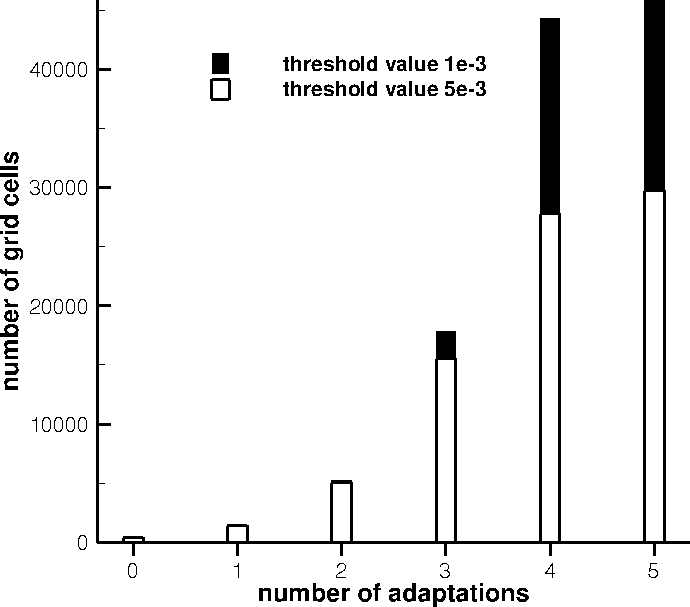}}
       \mbox{\includegraphics [height = 0.5\textwidth,clip]{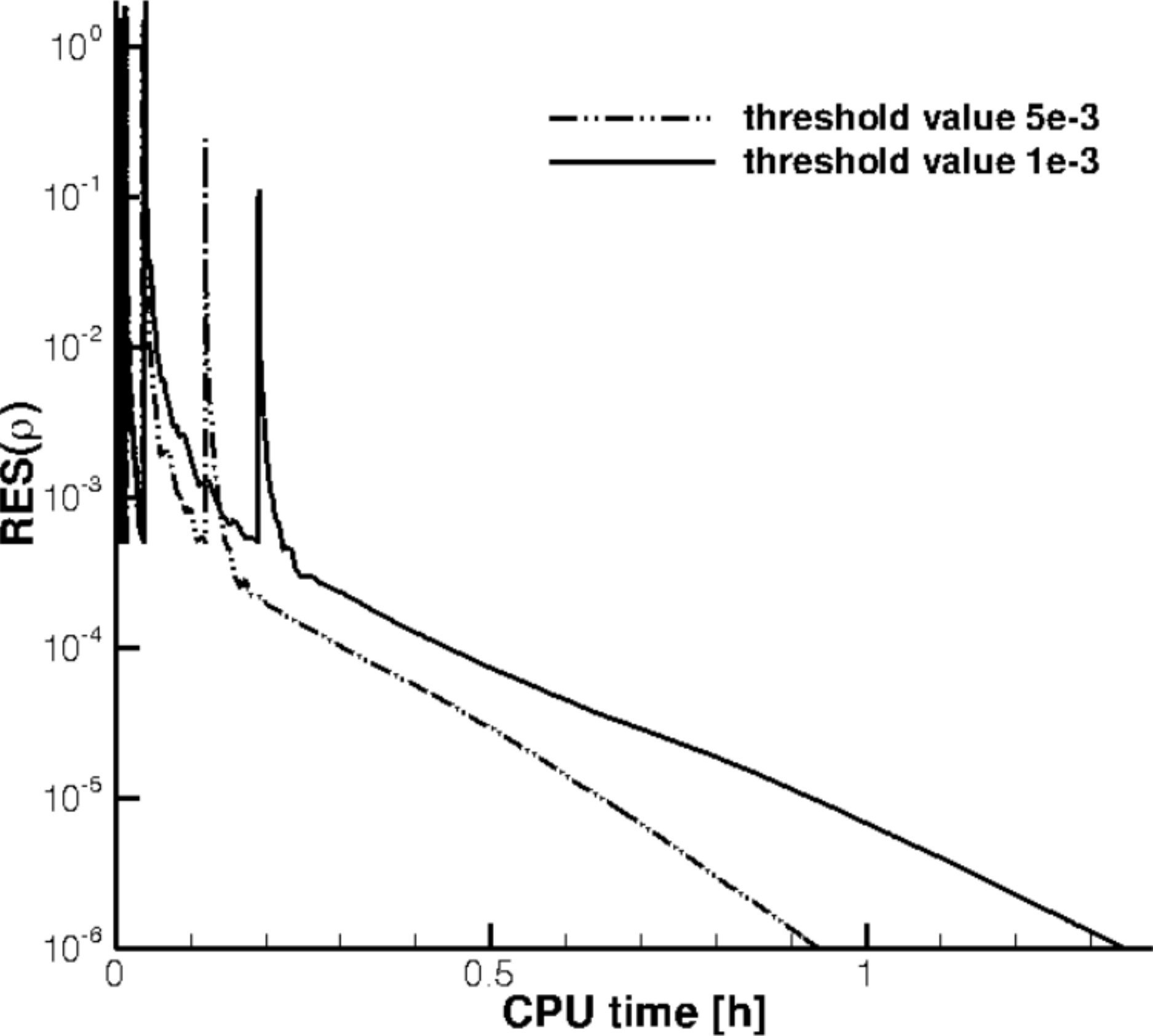}} }
              \caption{Evolution of grid cells (left) and computational time (right) over a compression corner at 15 degrees. Comparison between adaptive grids using two threshold values $\varepsilon_{thres}$=10$^{-3}$ and $\varepsilon_{thres}$=5$\times$10$^{-3}$. The highest refinement level is L=5. Inflow conditions: Re= 9.65$\times$10$^{6}$~1/m, M=6.35.}\label{2d-results-c}
  \end{figure*}
  
  \begin{figure*}[h!]
 \centerline{
    \mbox{\includegraphics [height = 0.5\textwidth,clip]{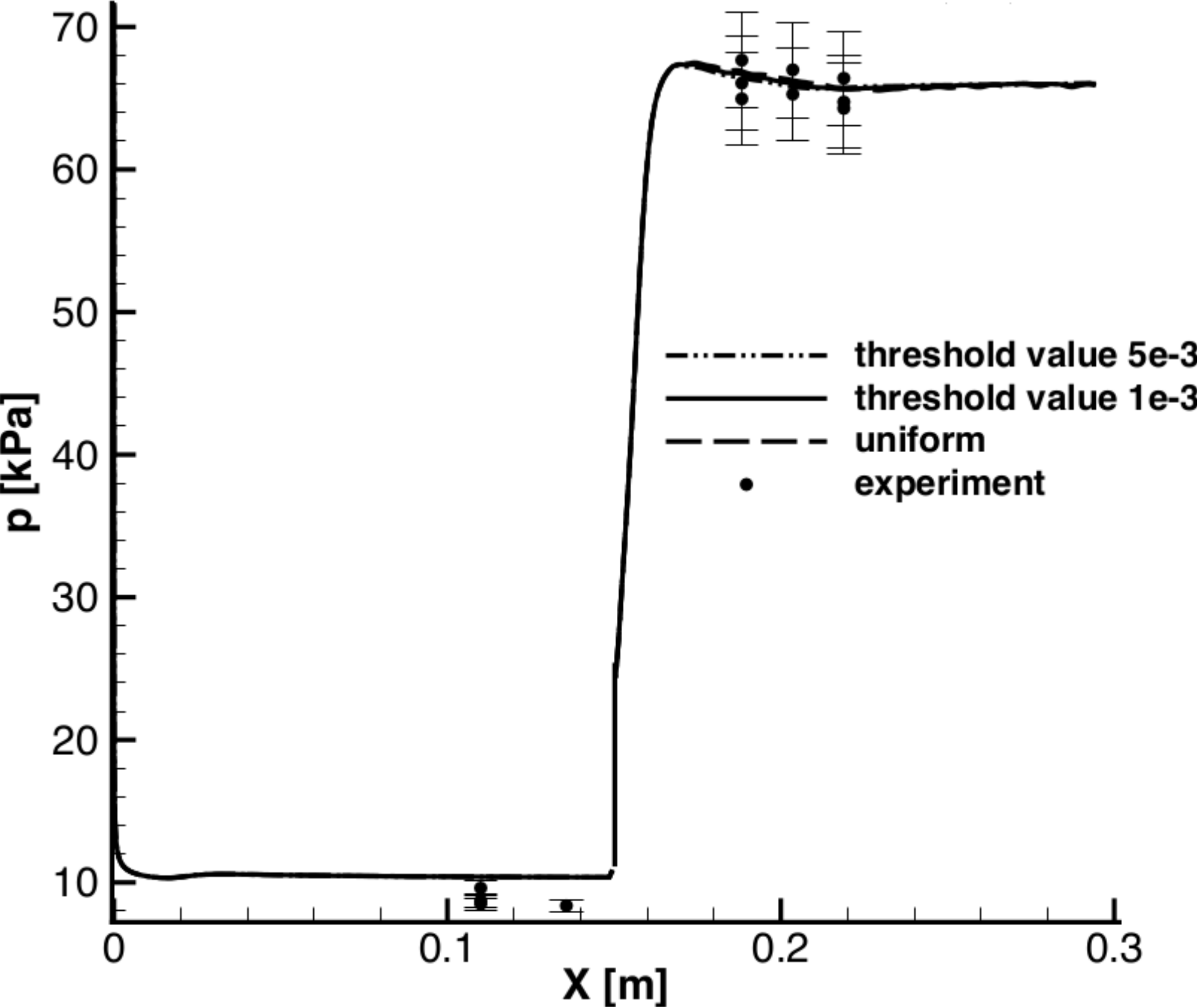}}
    \mbox{\includegraphics [height = 0.5\textwidth,clip]{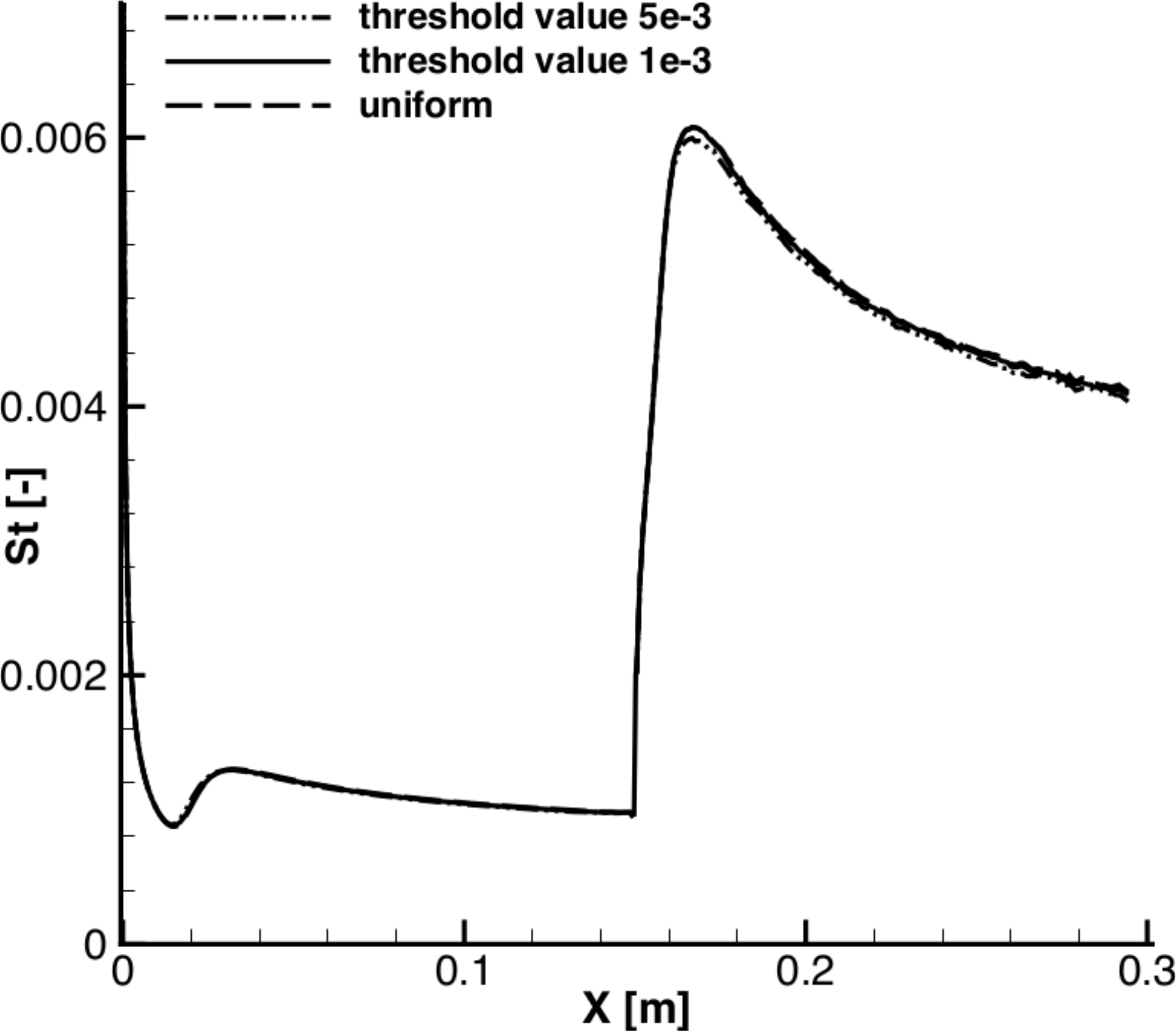}} }
        \caption{Pressure (left) and Stanton number (right) distribution over a compression corner at 15 degrees. Comparison between adaptive grids using two threshold values $\varepsilon_{thres}$=10$^{-3}$ and $\varepsilon_{thres}$=5$\times$10$^{-3}$ and uniform grid. The highest refinement level is L=5. Inflow conditions: Re= 9.65$\times$10$^6$~1/m, M= 6.35. Experimental results from \cite{Bosco:11}.}\label{2d-results-b}
          \end{figure*}
         
For the adaptive procedure to be meaningful, the obtained results must have the same accuracy as those obtained on a uniformly refined grid. To illustrate this two adaptive computations with two different threshold values $\varepsilon_{thres}=10^{-3}$ and $\varepsilon_{thres}=5\times10^{-3}$ are performed. It should be kept in mind that for a higher threshold value more details are neglected and fewer cells are refined. Thus the grid size is smaller. This directly affects the required CPU time (Figure \ref{2d-results-c}). Therefore, it is important to choose the threshold value as high as possible. However the threshold value has to be low enough to maintain the accuracy of the results of the uniformly refined grid. Figure \ref{2d-results-b} (right) shows that a threshold value of $\varepsilon_{thres}=10^{-3}$ is necessary to resolve the Stanton number accurately, whereas both adaptive computations resolve the pressure (Fig. \ref{2d-results-b} left) correctly.\\

\begin{figure*}[h!]
 \centerline{
    \mbox{\includegraphics [width = 0.6\textwidth,clip]{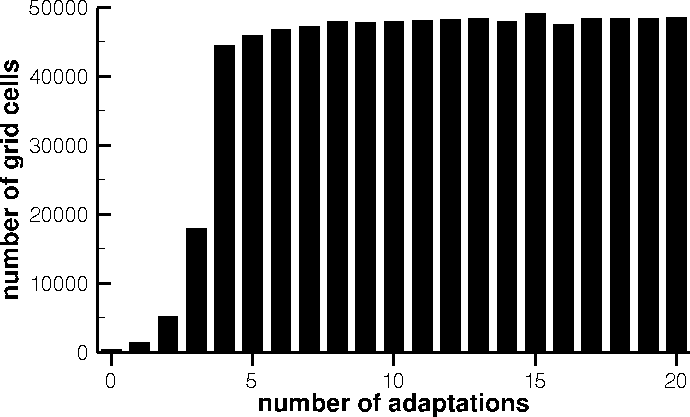}}}
        \caption{Evolution of cell numbers during the adaptive procedure for a two-dimensional compression ramp using five refinement levels and 20 adaptations.}\label{2d-moreadap}
  \end{figure*}

To show that the grid size converges to a constant value, an adaptive computation with 20 instead of 5 adaptations is performed. Note that the final 16 adaptations are performed on the highest refinement level $L=5$. Figure \ref{2d-moreadap} presents the evolution of grid cells during the adaptive procedure for this computation. During the first adaptations the grid size increases rapidly. With further adaptations the number of grid cells converges to a constant number. The adaptations on the highest refinement level modify the grid only slightly and do not improve the numerical solution. Thus, for all computations within this paper we perform only one additional adaptation on the highest refinement level to take the details on the highest refinement level into account.
\begin{figure*}[h!]
 \centerline{
  \mbox{\includegraphics [height= 0.45\textwidth,clip]{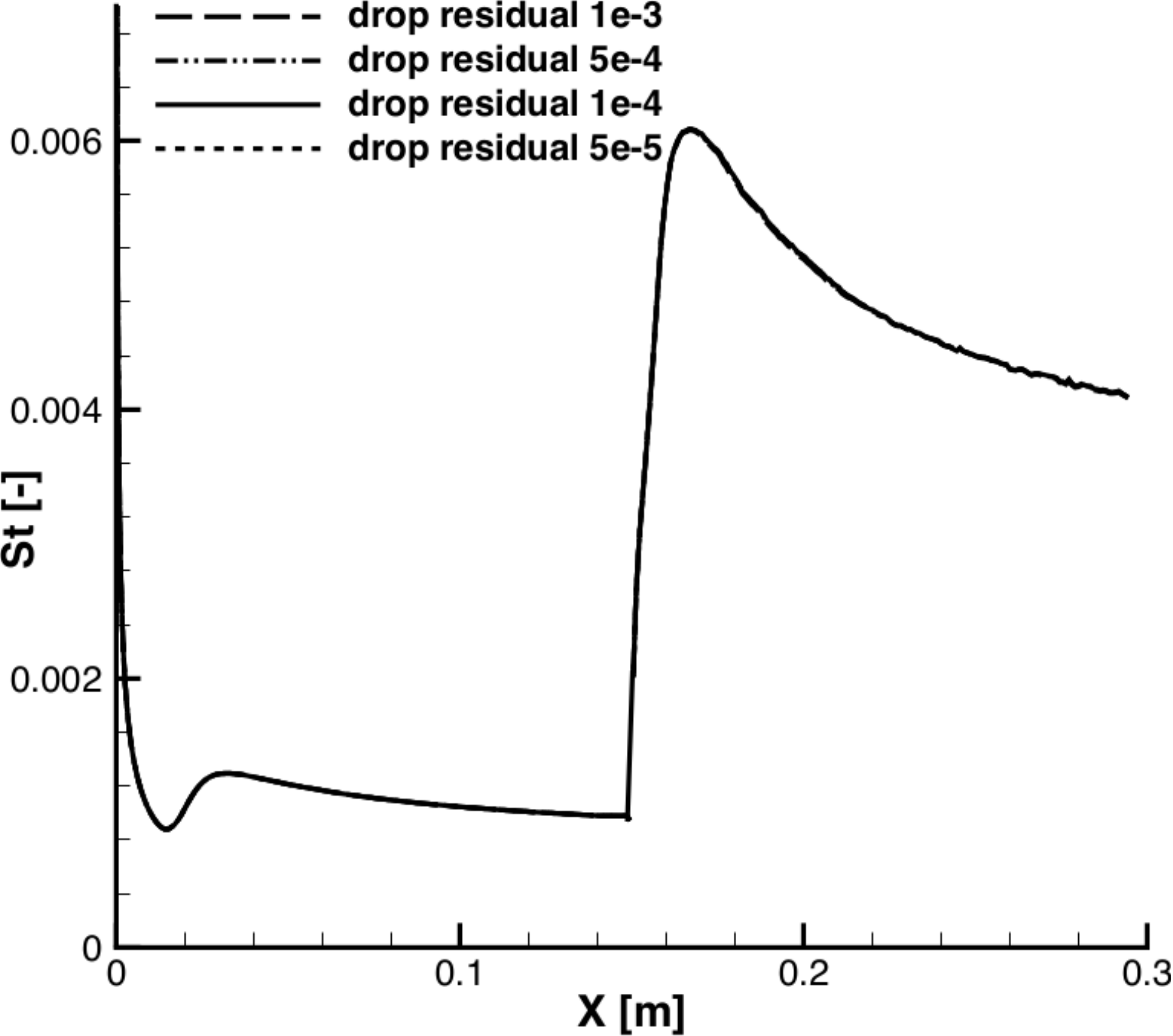}}
   \mbox{\includegraphics [height= 0.45\textwidth,clip]{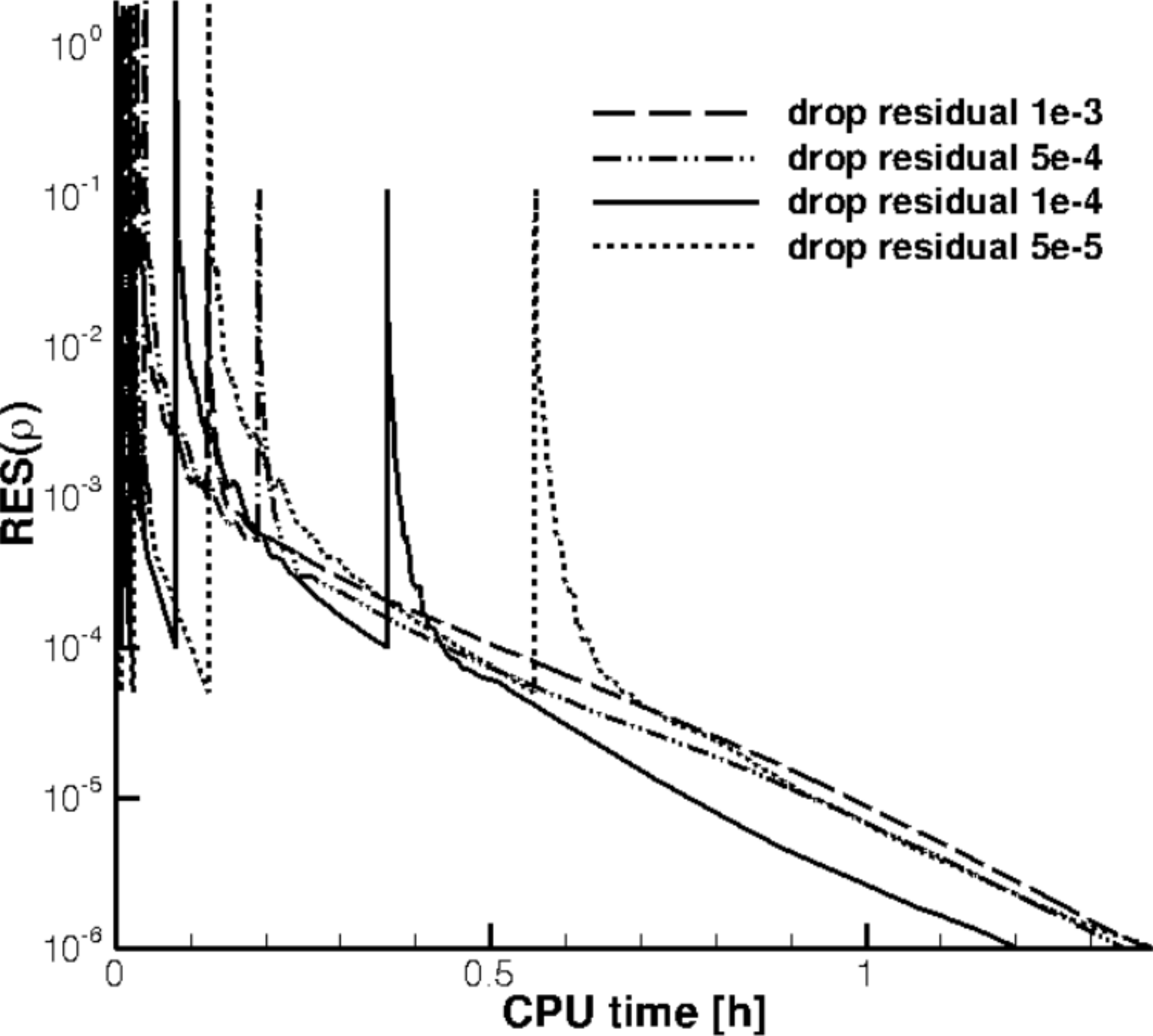}}
 }\caption{Stanton number distribution (right) and averaged density residual with respect to the computational time (left). Comparison between adaptive grids using different drop residuals. Inflow conditions: Re= 9.65$\times$10$^{6}$ 1/m, M= 6.35.}\label{drop-res}
  \end{figure*}
  
Next, we consider the influence of the drop residual at which the adaptations are performed. This parameter does not influence the solution accuracy (Fig. \ref{drop-res} (left)) but the CPU time of the computation (Fig. \ref{drop-res} (right)). In general, it is cheaper to compute a long time on the lower levels and start with a better initial guess on the next higher level. Thus the $\varepsilon_{drop}=10^{-3}$ computation takes longer than the $\varepsilon_{drop}=10^{-4}$ computation. But at some point this is not longer valid since the $\varepsilon_{drop}=5\times 10^{-5}$ computation requires more CPU time than the $\varepsilon_{drop}=10^{-4}$ computation. This implies that it makes no sense to fully converge the solution on the lower refinement level because the overall accuracy is too low. For this test case,  $\varepsilon_{drop}=10^{-4}$ is optimal. 
To illustrate the advantages of adaptive computations in contrast to classical non-adaptive computations, we compare the adaptive computation with the uniformly refined grid in terms of grid size, required iterations and CPU time. The uniformly refined grid is composed of 92,160 cells whereas the total number of cells necessary for the adaptive simulation is roughly half of that. Considering the characteristics of hypersonic grids, this gain is remarkable: Due to the supersonic speed of the flow, the grid starts directly at the leading edge of the ramp. In addition, the large thickness of the boundary layer, typical of the hypersonic flow regime, imposes the use of a fine grid in an ample portion of the domain adjacent to the solid surface, leaving apparently little room for grid coarsening. 
\begin{figure*}[h!]
 \centerline{
   \mbox{\includegraphics [height= 0.45\textwidth,clip]{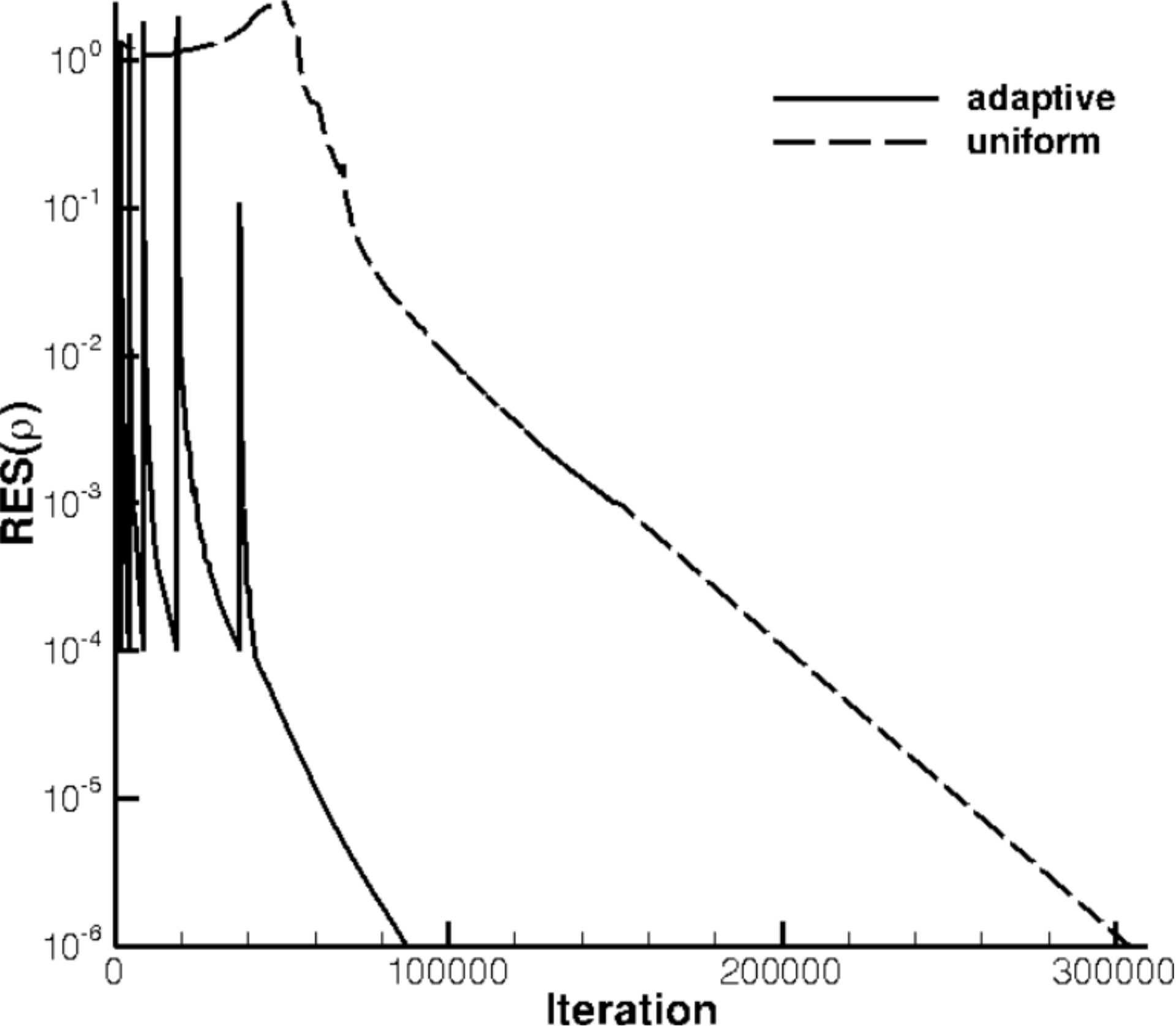}}
  \mbox{\includegraphics [height= 0.45\textwidth,clip]{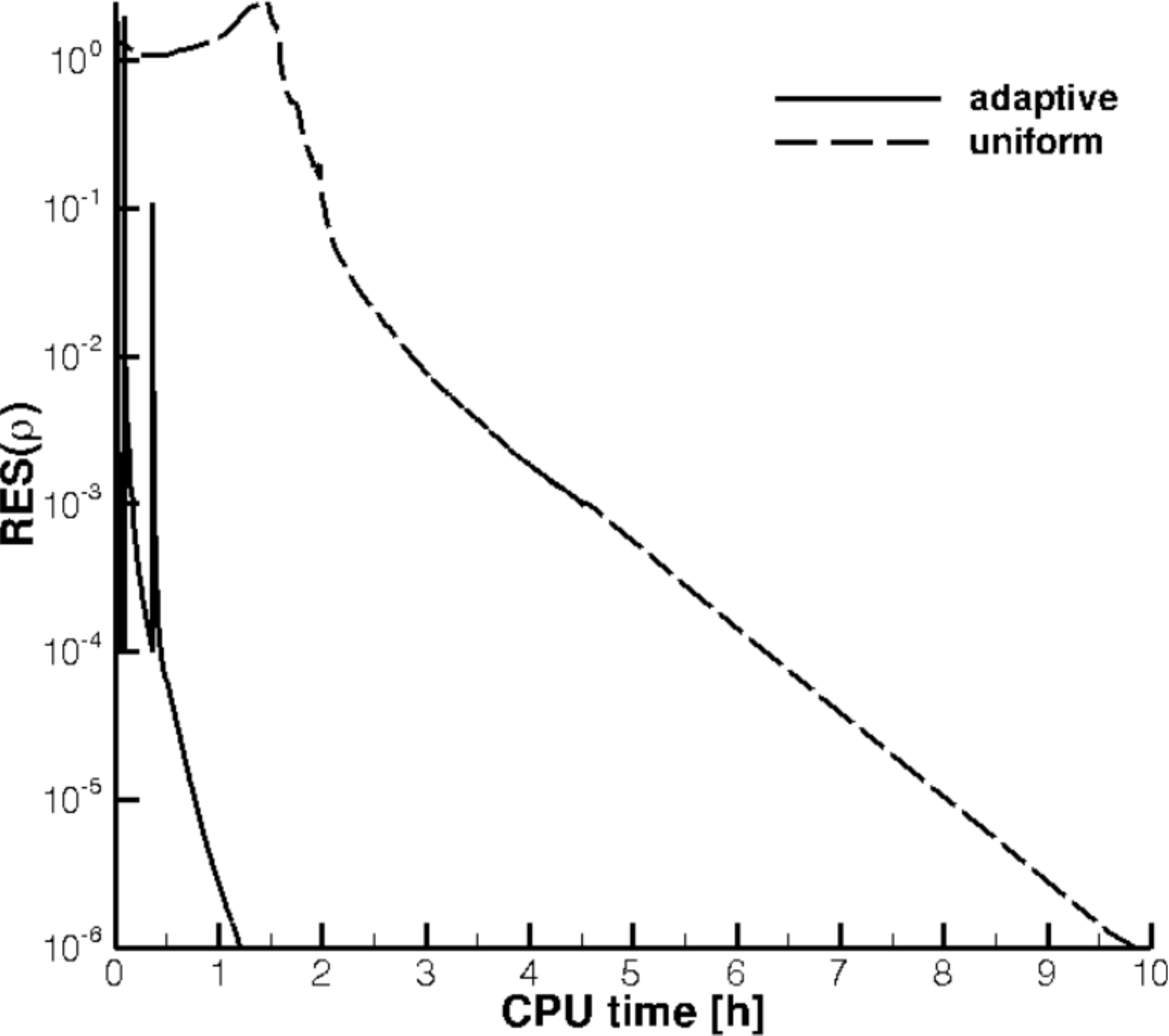}} }
        \caption{Averaged density residual with respect to the number of iterations (left) and computational time (right) for a compression corner at 15 degrees. Comparison between adaptive grid and uniform grid at L=5. Inflow conditions: Re= 9.65$\times$10$^{6}$~1/m, M= 6.35.}\label{convergence}
  \end{figure*}

In Figure \ref{convergence} (left), the normalized averaged density residual with respect to the number of iterations is shown. Here we see that by means of grid adaptation the number of iterations necessary to get a converged solution is one third (100,000 instead of 300,000 iterations). Since the computation starts on a coarse grid, the solution converges fast. During the adaptation procedure the new grid is initialized by means of the multiscale analysis of  the solution data of the old grid, which is a good initial guess for the finer grid and thus less iterations are necessary on the finer grids.
\\
Figure \ref{convergence} (right) shows the key advantage of using an adaptive strategy. Here the behavior of the normalized averaged density residual with respect to the computational time is shown. In case of an adaptive grid, the time required for completing the computation is one eighth of that necessary for the uniformly refined grid (1.5 hours instead of 10 hours). This is both due to the reduced number of iterations, as shown in Figure \ref{convergence} (left), and (mostly) to the reduced computational time per iteration due to the smaller grid size. 
In this way, as discussed before, the computational cost is strongly reduced without affecting the accuracy of the solution.
\subsection{Scramjet intake: two-dimensional results}

\begin{figure}[!ht]
\begin{center}
  \includegraphics[width=1\textwidth]{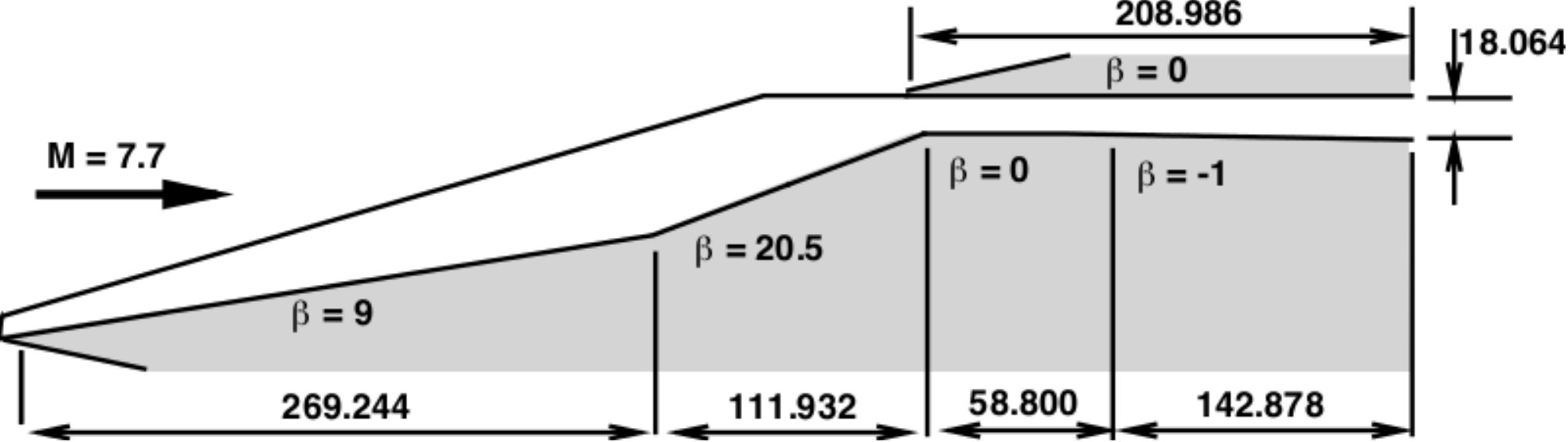}
  \caption{$xy$-plane of the considered scramjet geometry; dimensions are in millimeters.}
  \label{geo}
\end{center}
\end{figure}

In the next two sections, two- and three-dimensional results for a scramjet intake configuration are discussed. The intake model has been developed in the frame of the German Research Training Group GRK 1095 ``Aero-Thermodynamic Design of a Scramjet Engine for Future Space Transportation Systems'' \cite{Gaisbauer:07}. Figure \ref{geo} shows the geometry of the considered scramjet intake. The model has two exterior compression ramps and an interior section. The leading edge of the first ramp and the cowl lip are sharp. The model is 100 mm wide and has side walls on both sides. 

The configuration has been designed for an inflow Mach number $M_\infty=7.5$ and was tested at a slight off--design condition in the hypersonic shock tunnel facility TH2 in Aachen \cite{Neuen:06, Fischer}. The test conditions of the experimental campaign are listed in Table \ref{t:condi}. These values are used as inflow conditions in the simulations. During the experiments, pressure and heat transfer rate were measured by Kulite pressure probes and thermocouples, respectively. Since the transition point at the end of the first ramp is known from the experiments, the transition is modeled using a ``laminar box'' for this region. Within this ``laminar box'' the flow is forced to be laminar by setting the turbulent kinetic energy to zero.
\begin{table}[!h]
\caption{\label{t:condi}Test conditions for the scramjet intake configuration.}
\begin{ruledtabular}
\begin{tabular}{ccccc}
      $M_\infty$ [-] & $Re_\infty$ [$10^6$/m] &   $T_0$ [K]&   $T_\infty$ [K] &   $T_{\rm w}$ [K] \\\hline
            7.7  &  4.1            	     &  1520       &     125         &    300 		\\
 \end{tabular}
\end{ruledtabular}
\end{table}

\begin{figure*}[!ht]
\begin{center}
   \includegraphics[width=1\textwidth]{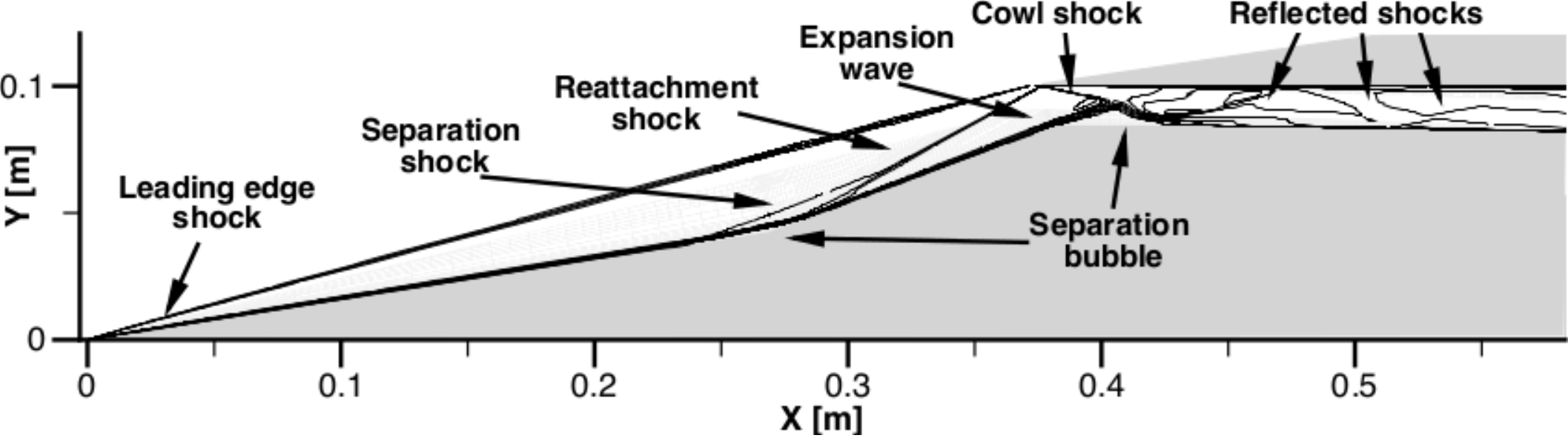}
   \caption{Two-dimensional adaptive computation of the scramjet intake showing the main physical flow phenomena via Mach number lines.}
   \label{Scramjet_Intake}	
\end{center}
\end{figure*}
The overall flow phenomena can be seen in Figure \ref{Scramjet_Intake}. The flow is first compressed through an oblique shock wave generated by the sharp leading edge. At the first ramp, a laminar boundary layer develops. In the kink between the first and second ramp, a small separation bubble is generated. Here the flow transitions from laminar to turbulent. 
Due to the off-design condition the reattachment shock hits the upper intake wall and deflects the oblique shock wave produced by the cowl lip slightly. The cowl shock interacts with the expansion fan and ramp boundary layer. Large adverse pressure gradients are produced by this interaction and cause a second separation bubble on the intake wall. In the interior region, the flow is going through several reflected shock waves.

The grid has 44 cells in the flow direction and 6 cells in the cross-flow direction on refinement level $L=0$.
To ensure a minimum wall distance of $10^{-6} $~m on the the final level $L=4$ grid, the grid points in wall-normal direction are stretched towards the walls using a Poisson distribution. Transverse to the wall the grid lines are almost always orthogonal to the walls to resolve the strong wall gradients accurately. 
\begin{table}
\caption{\label{adapt-2d-scramL4}Evolution of cell numbers during the adaptive procedure for a two-dimensional grid using four refinement levels.}
\begin{ruledtabular}
\begin{tabular}{ccccccc }
     refinement level   & $L$=0 & $L$=1  & $L$=2  &  $L$=3 & $L$=4 & $L$=4 final \\ \hline
        number of cells    & 264     &   1,056    & 3,813    & 14,136  &39,618 & 41,319 
 \end{tabular}
\end{ruledtabular}
\end{table}
The uniform grid on refinement level $L=4$ has 67,584 cells. Table \ref{adapt-2d-scramL4} shows the evolution of the grid size during the adaptive computation. In comparison to the uniformly refined grid, the adaptive grid is composed of only 60\% of the grid cells.

\begin{figure}[h!]
 \centerline{\includegraphics [width = 1\textwidth,clip]{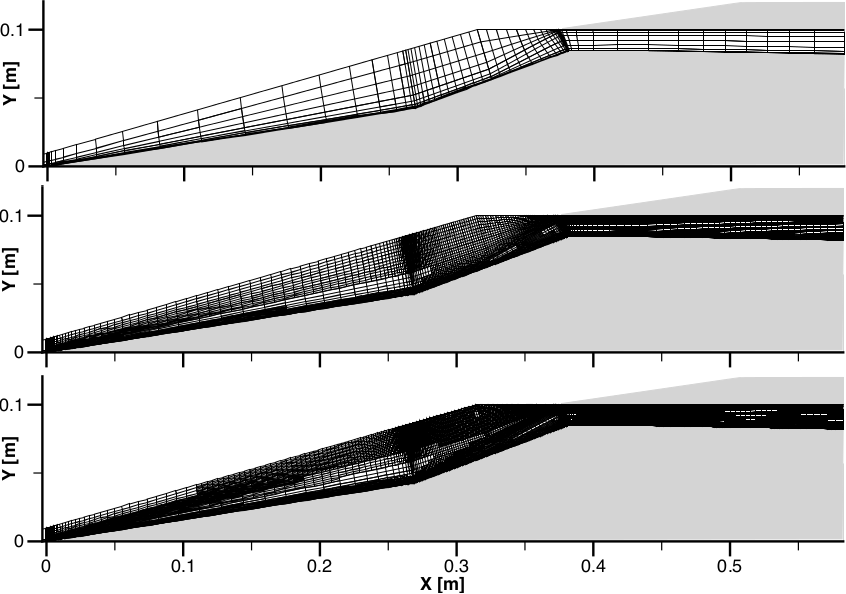}}
 \caption{Computational grids at different stages of the 2D computation. From top to bottom: initial grid at uniform refinement level L=1, intermediate adaptive grid at refinement level L=3, final adaptive grid at refinement level L=4.}\label{2d-grid-L1_scram}
\end{figure}
 \begin{figure}[h!]
 \centerline{\includegraphics [width = 1\textwidth,clip]{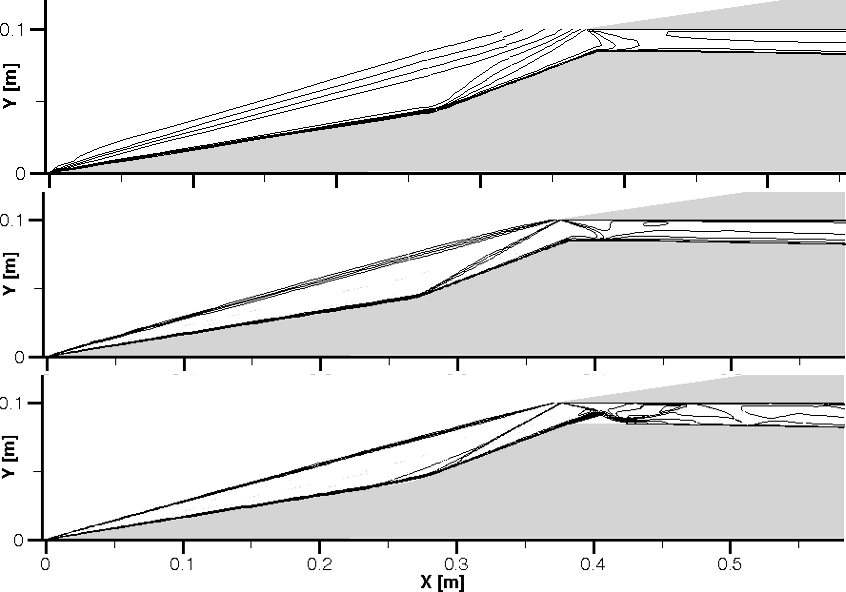}}
 \caption{Mach number contours at different stages of the 2D computation. From top to bottom: initial grid at uniform refinement level L=1, intermediate adaptive grid at refinement level L=3, final adaptive grid at refinement level L=4.}\label{2d-grid-L1_scramM}
\end{figure}

Figure \ref{2d-grid-L1_scram} presents the grid at the different refinement levels and Figure \ref{2d-grid-L1_scramM} the Mach number contours corresponding to the different grid levels. During the first and the second adaptation, most cells are refined. The third adaptation only refines the cells near the shock waves and in the boundary layers as well as the separation areas. After the third adaptation, the grid is on the highest refinement level. During the fourth adaptation, only a reordering of the cells is performed. Hence, the number of cells stays nearly the same after the third and fourth adaptation.\\
\begin{figure*}[h!]
 \centerline{
 \mbox{\includegraphics [width = 0.5\textwidth,clip]{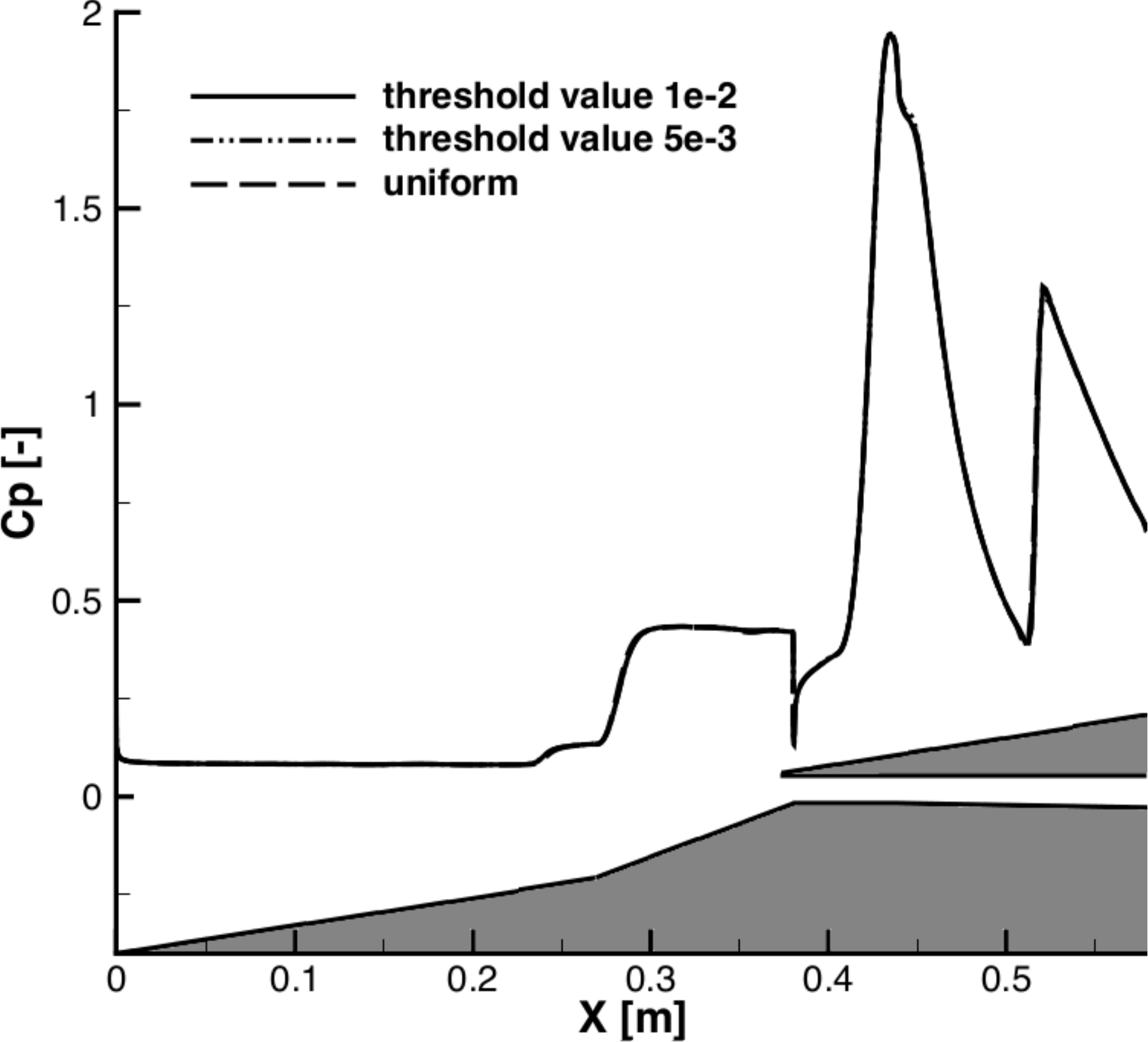}}
 \mbox{\includegraphics [width = 0.5\textwidth,clip]{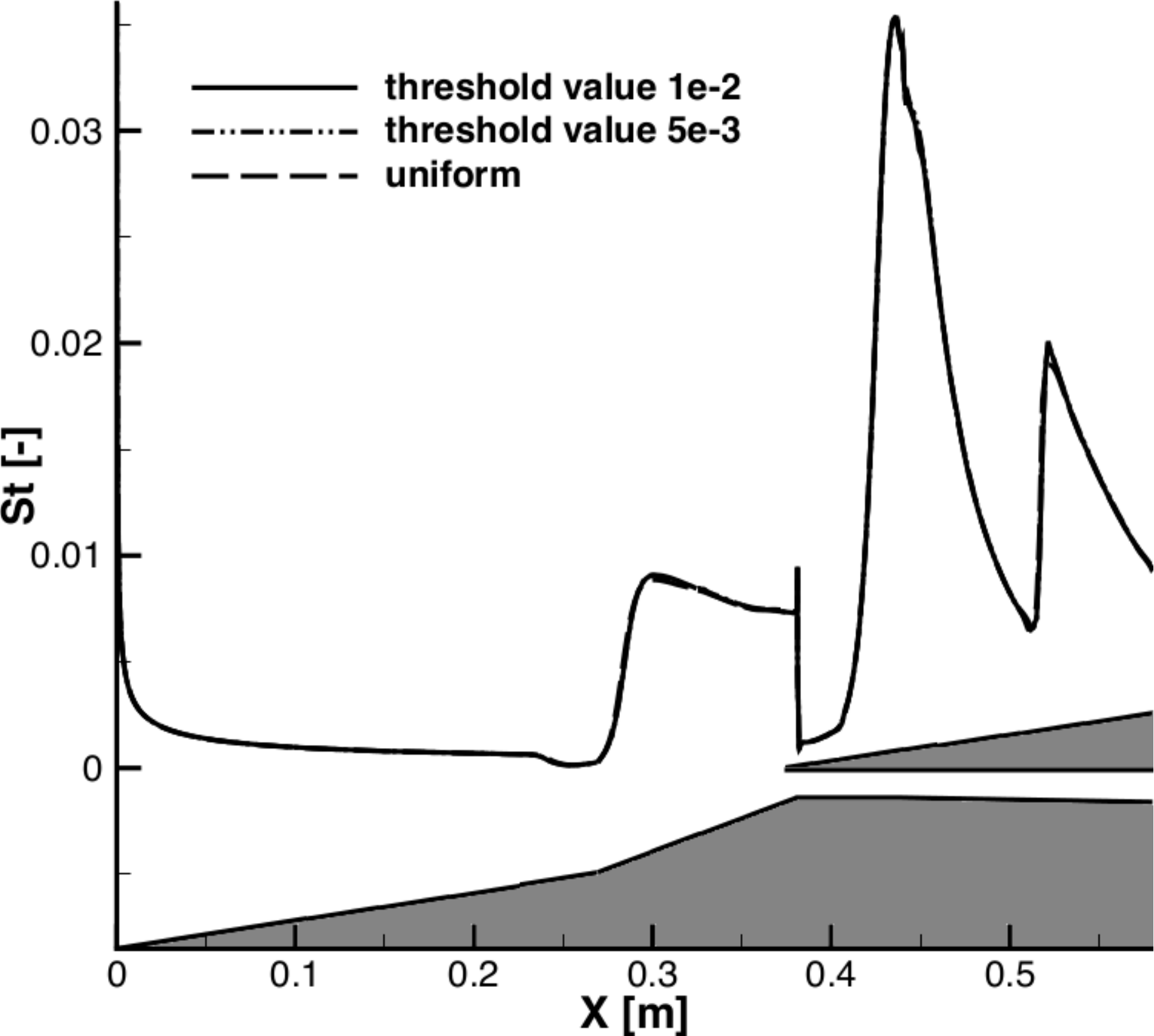}} }
 \caption{Pressure coefficient (left) and Stanton number (right) distribution at lower wall of the two-dimensional scramjet intake. Comparison between two adaptive grids using different threshold values and a uniform grid. }\label{2d-wall-scram-a}
\end{figure*} 

Figure \ref{2d-wall-scram-a} shows the pressure coefficient, 
\begin{equation}
c_p = \frac{p-p_{\infty}}{\frac{1}{2}\rho_{\infty}  u_{\infty}^2} ,
\end{equation}

\noindent and the Stanton number (\ref{stanton}) at the lower intake wall for two different adaptive level $L=4$ computation using different threshold values and the uniformly refined $L=4$ grid. 
The geometry of the intake is also shown. First, the general behavior of the quantities are discussed. Then the choice of the threshold value for the adaptive procedure is analyzed. 
    
Most of the flow features shown in Figure \ref{Scramjet_Intake} can also be identified in the plot of the pressure coefficient and the Stanton number. Along the first ramp, the flow is laminar and the pressure and Stanton number are low. At the end of the first ramp, the separation shock occurs and thus the pressure rises. The separation bubble is clearly visible in the pressure plateau. In this region, the flow is still laminar and therefore the Stanton number drops. The reattachment shock produces the next increase of the pressure and the Stanton number. Along the second ramp, both quantities remain nearly constant. At the end of the ramp, the expansion occurs and thus the pressure and Stanton number start to decrease. When entering the interior region the flow separates due to the impinging cowl shock wave and interacts with the expansion fan. Hence, the pressure rises directly instead of remaining at the lower level. The flow is fully turbulent here which can be seen in the increase of the Stanton number over the separation region. The reattachment shock of the second separation yields strong compression and intense heating visible as peaks in both, pressure and Stanton number. This shock is reflected at the upper intake wall and hits the lower intake wall again. Due to this shock reflection, the second peak of the pressure and the Stanton number occurs.

For this test case a threshold value of $\varepsilon_{thres}=10^{-2}$ is sufficient since the obtained results show no significant difference to the results obtained on the uniformly refined grid. 

To compare the performance and accuracy of the adaptive simulation, a simulation on a structured grid that was specifically designed and optimized for this intake is performed \cite{Nguyen:10}. It contains 211,968 cells (1,104 points in flow direction and 192 points in wall-normal direction). The grid points in the wall normal direction are stretched by a Poisson distribution in order to achieve a minimum wall distance of $2\times10^{-6}$~m. The grid lines transverse to the wall are almost perpendicular to the walls. Nguyen \cite{Nguyen:10} performed a grid convergence study using a coarser grid (minimum wall distance of $4\times10^{-6}$~m, approximately 100,000 cells) and a finer grid (minimum wall distance of $1\times10^{-6}$~m, approximately 400,000 cells). It was shown that the results of the medium grid can be considered as grid-converged. Thus, this grid is used for comparison with the adaptive computation.

\begin{figure*}[h!]
 \centerline{
   \mbox{\includegraphics [width = 0.5\textwidth,clip]{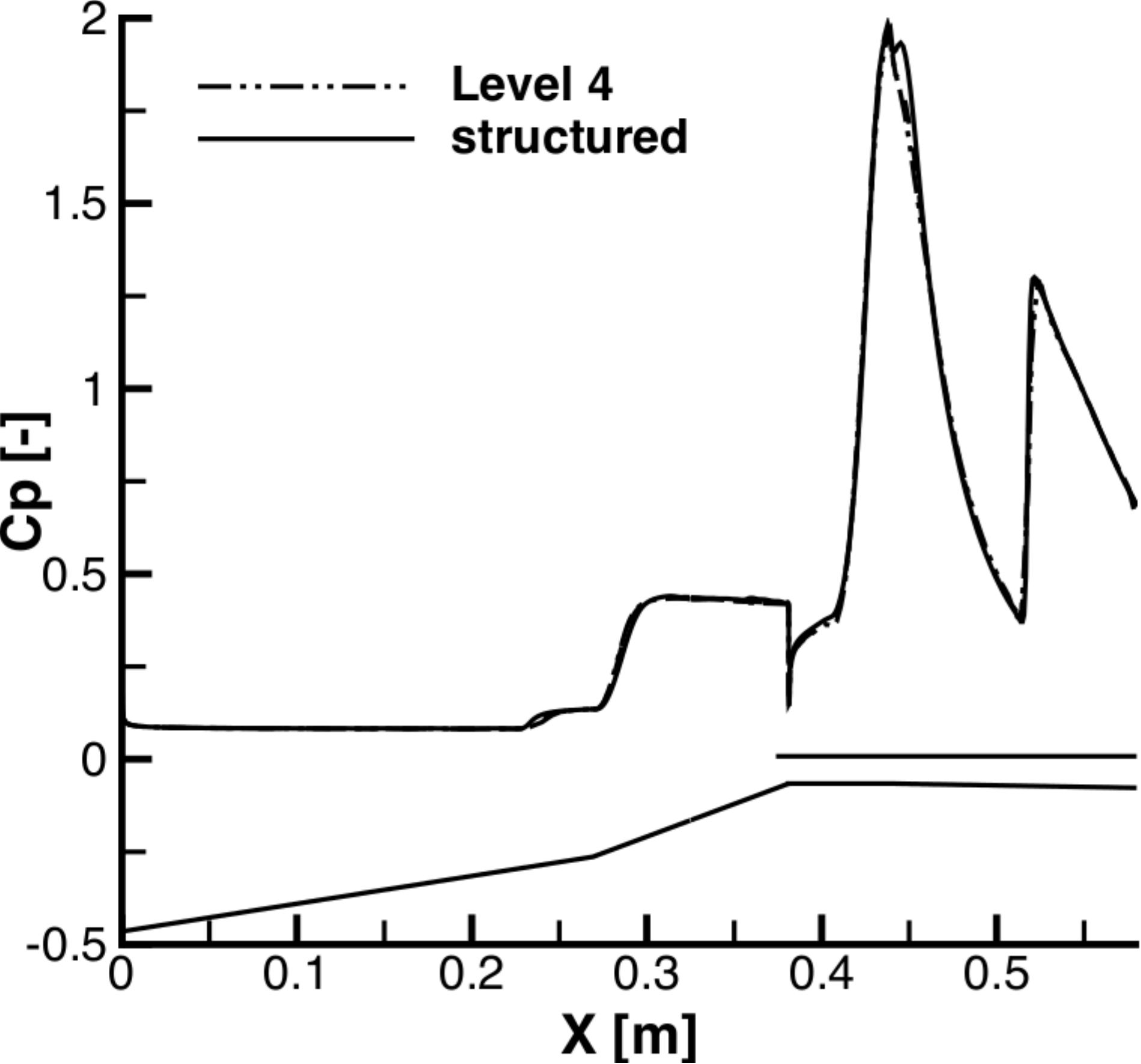}}
    \mbox{\includegraphics [width = 0.5\textwidth,clip]{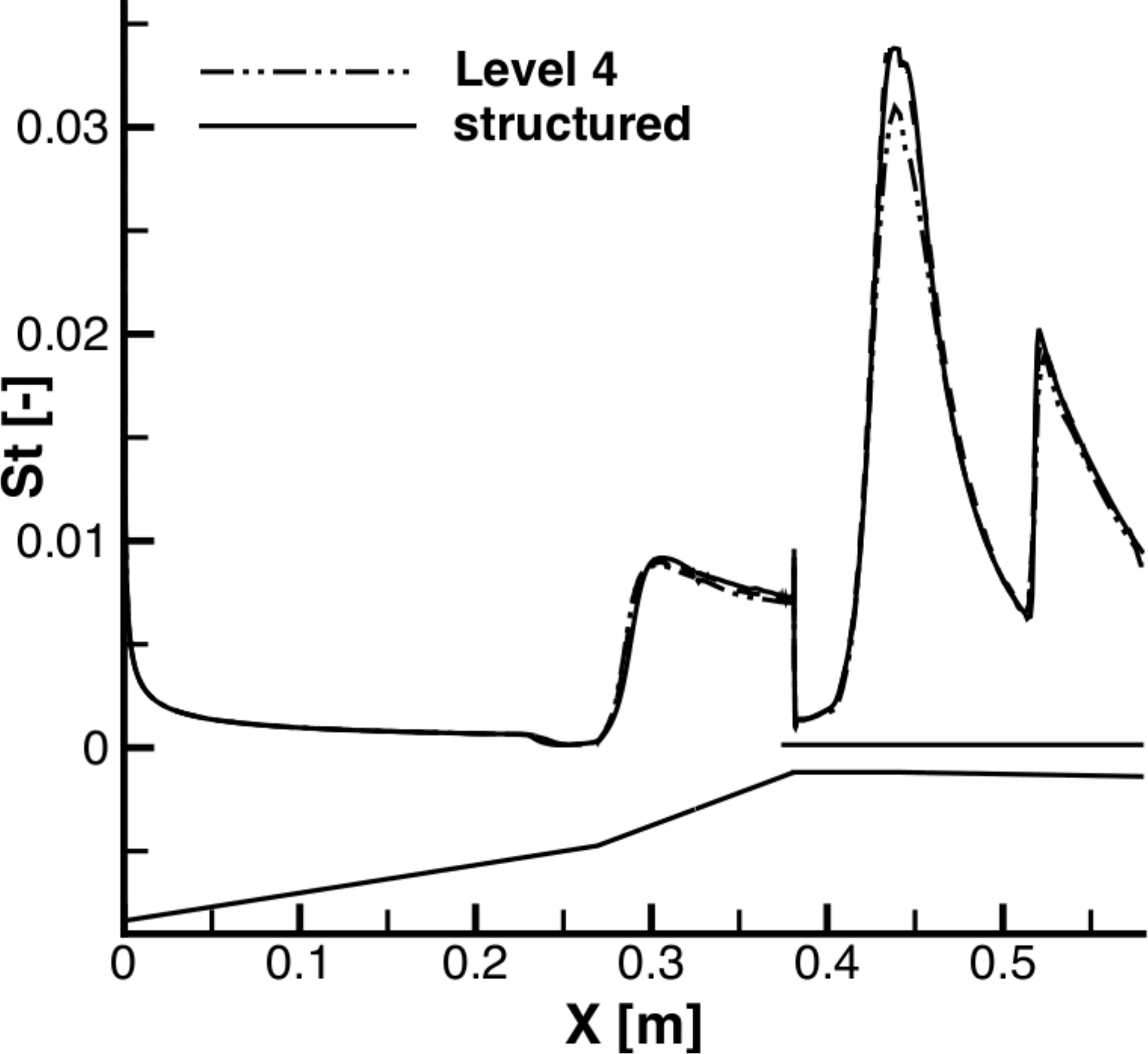}} }
 \caption{Pressure coefficient (left) and Stanton number (right) distribution at lower wall of the two-dimensional scramjet intake. Comparison between the adaptive grid on refinement level L=4 and a non-adaptive, structured grid. }\label{2d-wall-scram-b}
  \end{figure*} 
  
\begin{figure*}[h!]
 \centerline{
   \mbox{\includegraphics [width = 0.5\textwidth,clip]{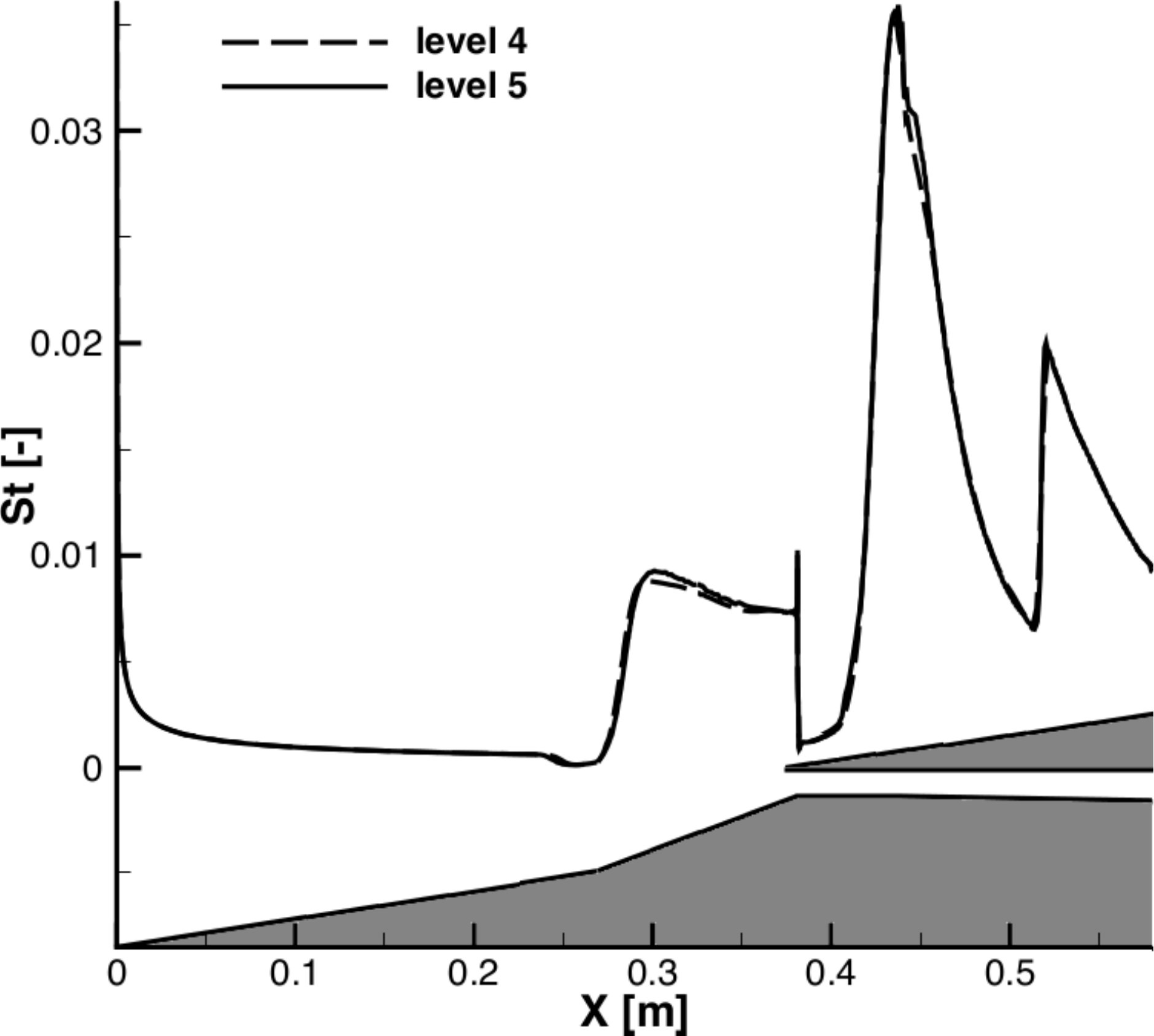}}
   }
 \caption{Stanton number distribution at lower wall of the two-dimensional scramjet intake. Comparison between two adaptive grids using 4 and 5 refinement levels. }\label{2d-grid_conv-scram}
  \end{figure*}

Comparing the wall distributions for the adaptive and the structured grid simulations, the overall agreement is very good except for the reattachment peak of the second separation ($x \approx 0.45$ m). Here, a small difference in the pressure coefficient and a larger discrepancy in the Stanton number is visible. Indeed, this is a difficult flow region because of the strong compression of streamlines close to the wall. Nguyen et al.~\cite{Nguyen:10} did not obtain grid convergence for their structured grid in this area but had to use the shown grid due to resource constrains, being aware of an underprediction of the peak heating.
Regarding the grid convergence of the adaptive computation, Figure \ref{2d-grid_conv-scram} shows the Stanton number distribution of the lower intake wall for the $L=4$ and $L=5$ computations. Here, the two computations are hard to distinguish;
this agreement confirms that the adaptation is able to detect and resolve the strong gradient correctly. 

\begin{figure*}[h!]
 \centerline{\includegraphics [width =1\textwidth,clip]{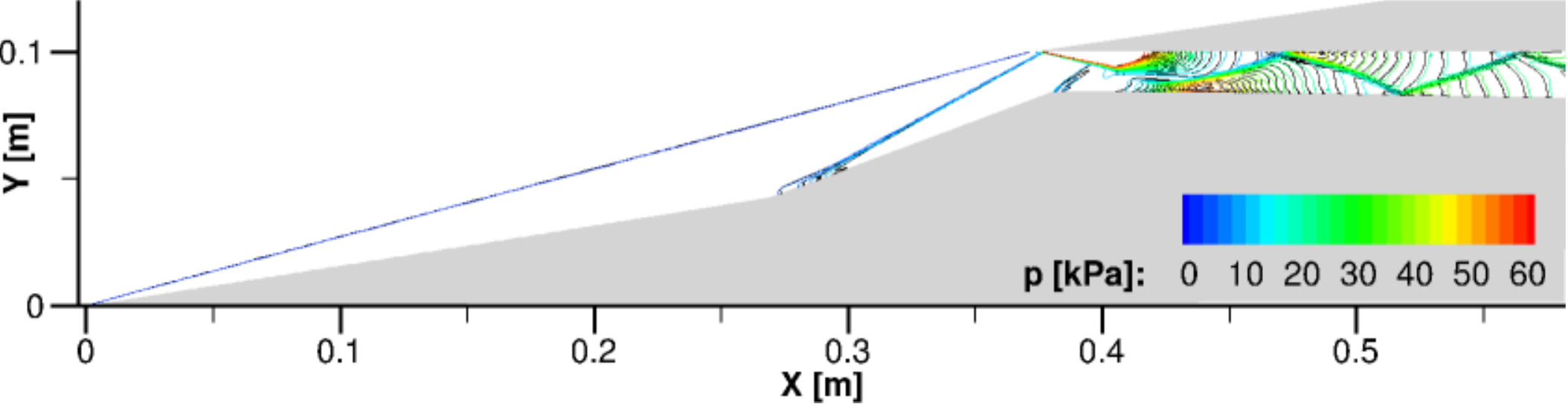} }
\caption{Pressure lines for the adaptive computation (colored) and the non-adaptive computation (black) of the two-dimensional scramjet intake.}\label{2d-p-scram}
  \end{figure*}

For comparing the flow field of the non-adaptive and adaptive computation, Figure \ref{2d-p-scram} shows the pressure distribution for both computations.

 \begin{figure*}[h!]
 \centerline{\includegraphics [width =0.5\textwidth,clip]{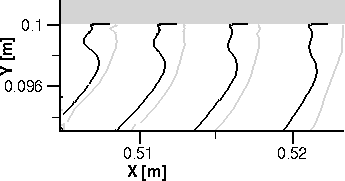} }
        \caption{Pressure lines at the top wall of the interior part. Comparision of the adaptive computation (grey lines) and the non-adaptive computation (black lines) of the two-dimensional scramjet intake. }\label{2d-p-zoom}
  \end{figure*} 
  
Although the adaptive grid consists only of 41,000 cells ($1/5$ of the structured grid), only small differences from the non-adaptive computation can be seen. Close to the wall, the adaptive computation resolves the gradients even better due to the smaller first wall distance. Figure \ref{2d-p-zoom} shows the pressure lines close to the top wall for the non-adaptive computation (black lines) and the adaptive computation (grey lines). Since the pressure is constant in the boundary layer, the pressure lines of the adaptive computation are more physical. Hence, close to the wall the adaptive computation is more accurate than the non-adaptive one. In the interior, the adaptive computation is as accurate as the non-adaptive computation. 

\begin{figure*}[h!]
 \centerline{
    \mbox{\includegraphics [height = 0.47\textwidth,clip]{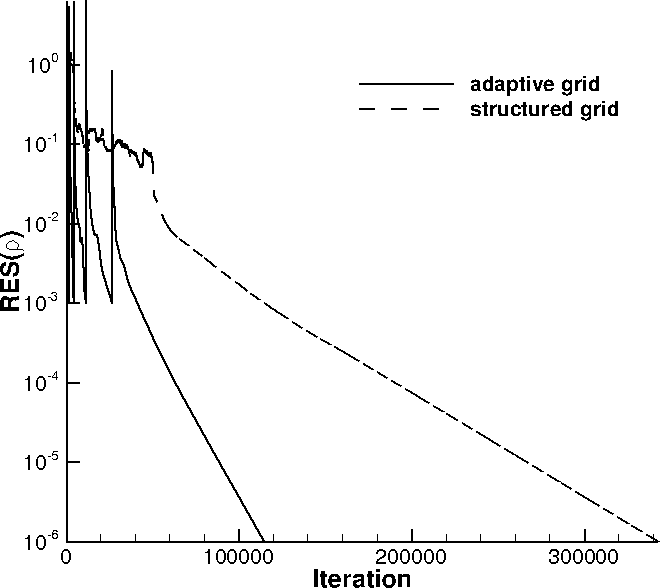}}
    \mbox{\includegraphics [height = 0.47\textwidth,clip]{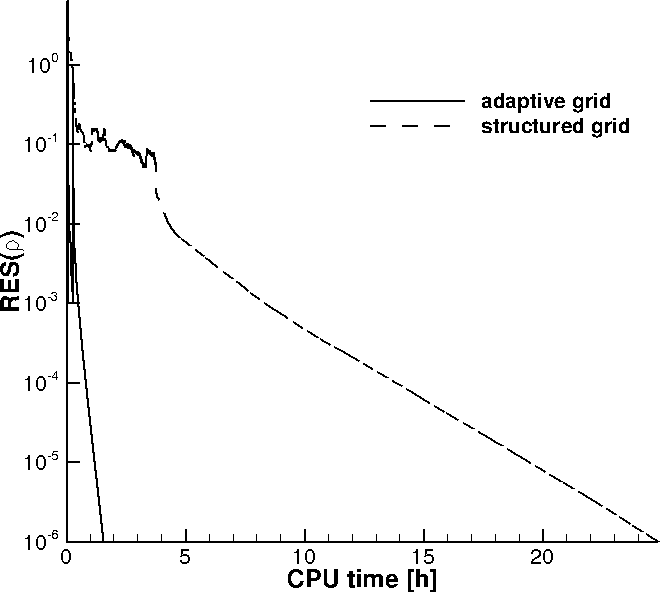}} }
        \caption{Behavior of the averaged density residual with respect to the number of iterations (left) and computational time (right) for a two-dimensional scramjet intake. Comparison between adaptive grid and non-adaptive structured grid. Inflow conditions: Re=  4.1$\times$10$^{6}$~1/m, M= 7.7.}\label{convergence_2d-scram}
  \end{figure*}

After proving the correctness of the adaptive results, we compare the performance of the adaptive simulation to the non-adaptive simulation on the structured grid. Figure \ref{convergence_2d-scram} shows the residual drop with respect to number of iterations and CPU time. Since the adaptive grid is only locally and where necessary refined, it has only half of the grid cells of the uniform grid at the highest refinement level. In comparison to the structured non-adaptive grid, which was specially designed and optimized for this application, the adaptive grid has only one fifth of the cells. 
The adaptive computation only runs for around 100,000 iterations whereas the non-adaptive computation requires more than 300,000 iterations. Due to the different grid sizes, performing one iteration on the adaptive grid (41,000 cells) requires less CPU time than performing one iteration on the non-adaptive grid (210,000 cells). Hence the adaptive computation only needs 1.5 hour, while the non-adaptive computation runs for 25 hours. 

This is a significant savings in CPU time without losing accuracy of the results. In addition, in contrast to the adaptive grid a lot of work and expertise was necessary to produce and optimize the non-adaptive, structured grid. 
Hence adaptive computation can fasten the grid generation process before the computation as well as shorten the computation time due to smaller grid sizes. 

\subsection{Scramjet intake: three-dimensional results}
To take three-dimensional effects into account, the two-dimensional grid described above has been extruded in the $z$-direction for half of the intake width. On the coarsest level, five cells have been stretched towards the side walls in the $z$-direction, using again a Poisson distribution to achieve a minimum wall distance of $10^{-6}$~m on level $L=4$. The point distribution in the interior part differs slightly from the two-dimensional grid. The evolution of the grid size during the adaptive procedure is shown in Table \ref{adapt-3d-scramL4}. The uniform grid on the highest refinement level is composed of 4,669,440 cells.

Due to the computational effort we did not analyze the grid convergence in 3D by computing a uniformly refined grid on Level $L = 5$. Since we showed grid convergence in 2D for the same test case using the same numerical parameters and a similar grid, we assume the resolution of the 3D computation to be sufficient. Instead we will be showing solution accuracy by comparing to results of a uniformly refined grid on the same level , i.e., $L = 4$.

\begin{figure*}[h!]
 \centerline{\includegraphics [width = 0.5\textwidth,clip]{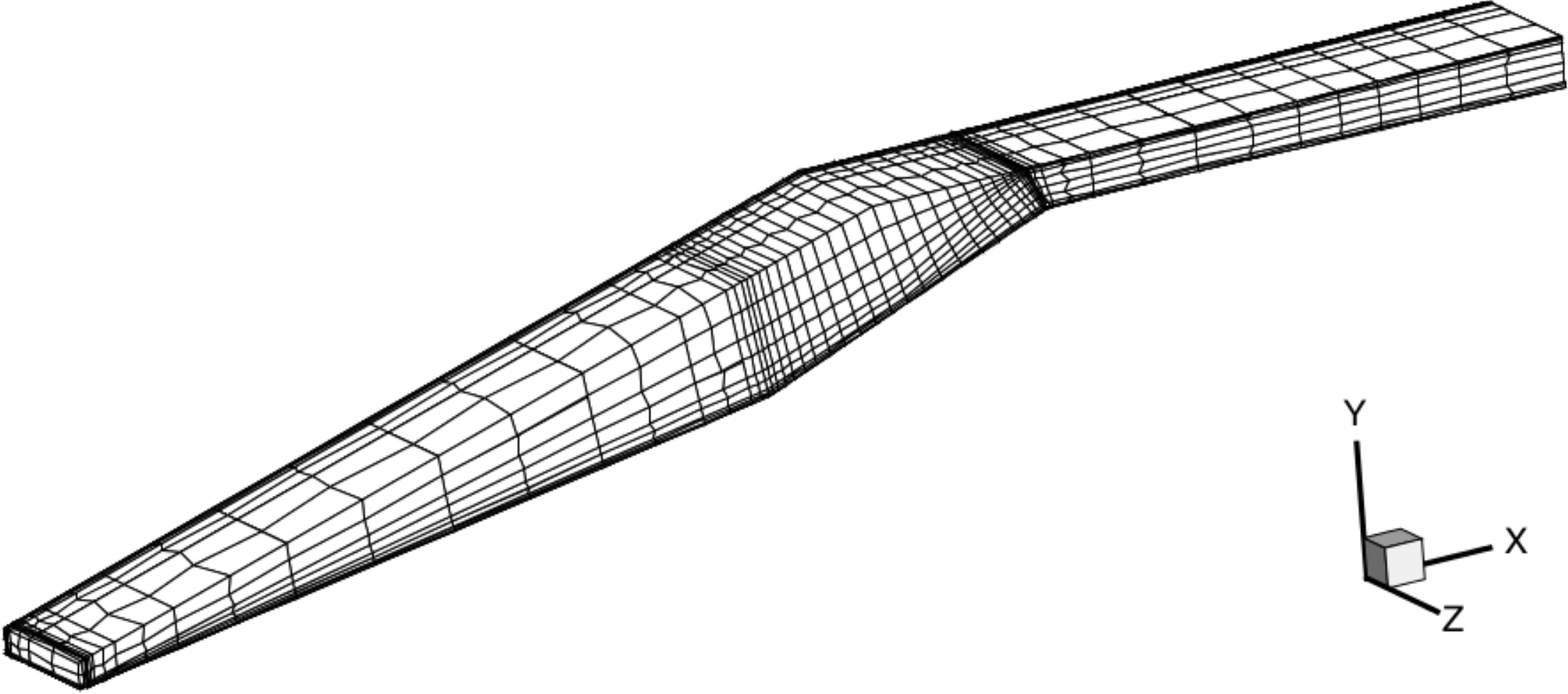}}
        \caption{Initial grid at uniform refinement level L=1.}\label{3d-grid-initial}
  \end{figure*}

\begin{figure*}[h!]
 \centerline{\includegraphics [width = 0.5\textwidth,clip]{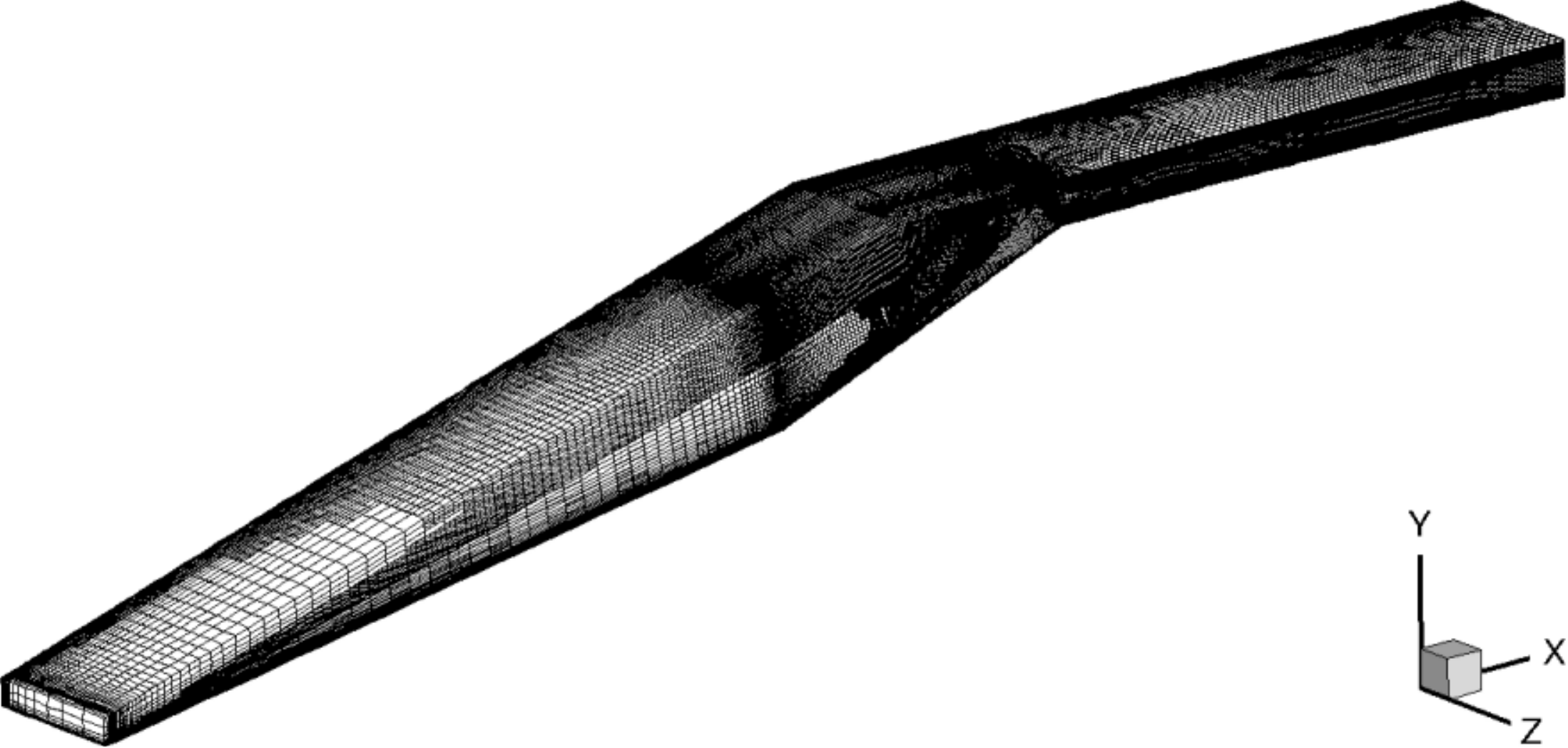}
  \includegraphics [width = 0.5\textwidth,clip]{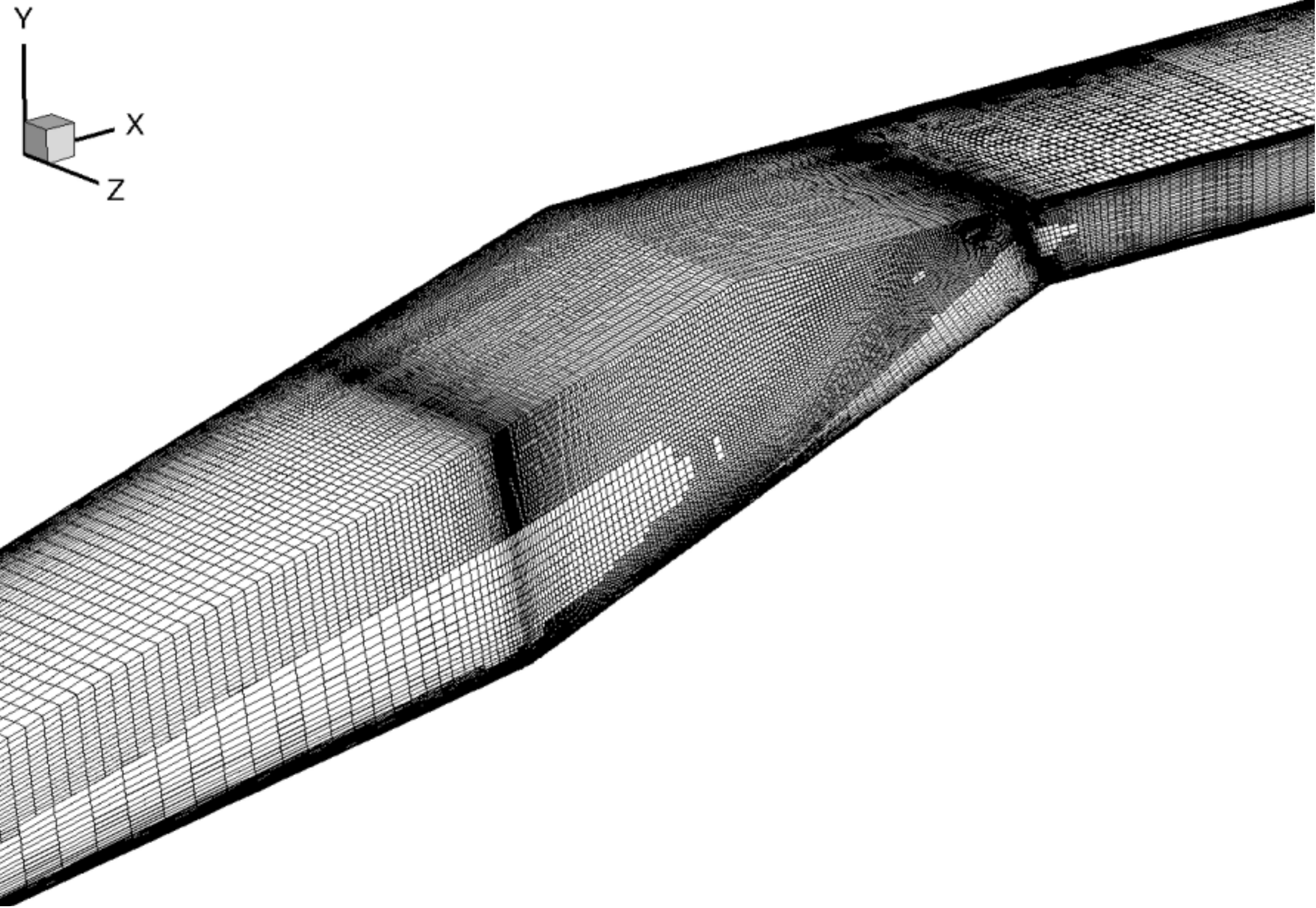}}
        \caption{Final computational grid at refinement level L=4 after the last adaptation.}\label{3d-grid-final}
  \end{figure*}

 \begin{table}[h]
\caption{\label{adapt-3d-scramL4}Evolution of cell numbers during the adaptive procedure for a three-dimensional grid using four refinement levels.}
\begin{ruledtabular}
\begin{tabular}{ccccccc}
 refinement level   & $L$=0 & $L$=1  & $L$=2  &  $L$=3 & $L$=4 & $L$=4 final\\ \hline
   number of cells   & 1,140  &   9,120& 72,960 & 550,696  &3,069,527  & 3,422,208  \\
 \end{tabular}
\end{ruledtabular}
\end{table}
Figures \ref{3d-grid-initial} and \ref{3d-grid-final} show the uniform initial grid at level $L=1$ and the final, locally adapted grid at level $L=4$. As in two dimensions, the final adapted grid is refined in important areas such as shock waves, boundary layers and shear layers due to separation. Elsewhere, the grid cells are on a coarser level.


 \begin{figure*}[h!]
 \centerline{
   \includegraphics [width = 0.5\textwidth,clip]{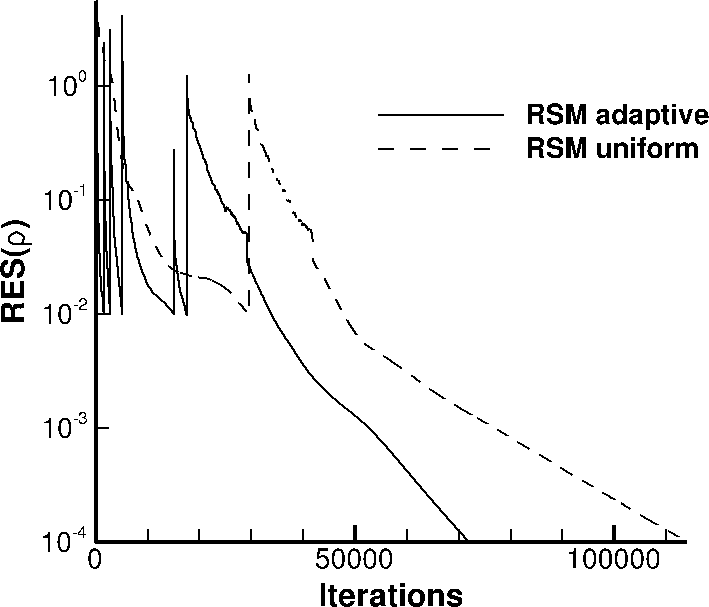}
    \includegraphics [width = 0.5\textwidth,clip]{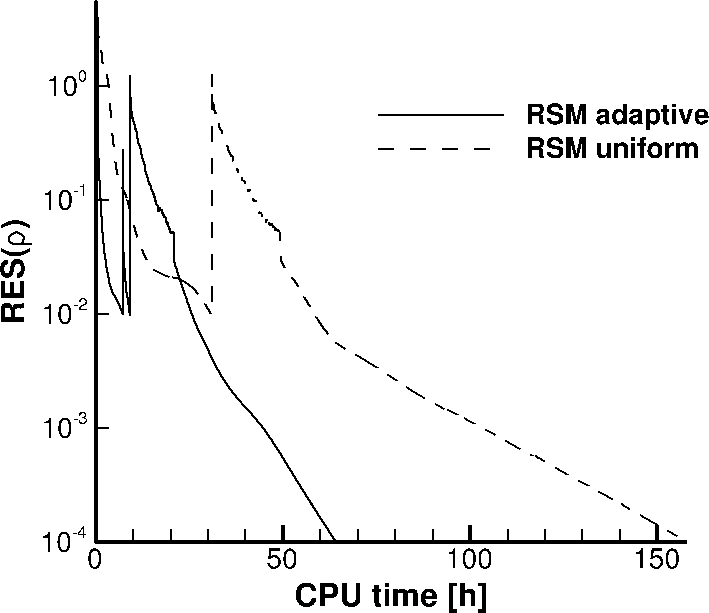}
    }
 \caption{Behavior of the averaged density residual with respect to the number of iterations (left) and computational time (right) for a three-dimensional scramjet intake. Comparison between adaptive grid and uniform grid at L=4. Inflow conditions: Re= 4.1$\times$10$^{6}$ 1/m, M= 7.7.}\label{convergence_3d-scram}
  \end{figure*}
  
First of all, the performance of the simulation using the adaptive procedure is compared with the simulation performed on a uniformly refined grid at the highest refinement level $L=4$. The averaged density residual drop with respect to the number of iterations and CPU time is shown in Figure \ref{convergence_3d-scram}.
The adaptive procedure requires two third the number of iterations needed for a uniform grid and, due to the smaller size of the final grid, the computational time can be decreased to around one third with respect to the uniform grid. 

    \begin{figure*}[h!]
 \centerline{
    \mbox{\includegraphics [width = 0.5\textwidth,clip]{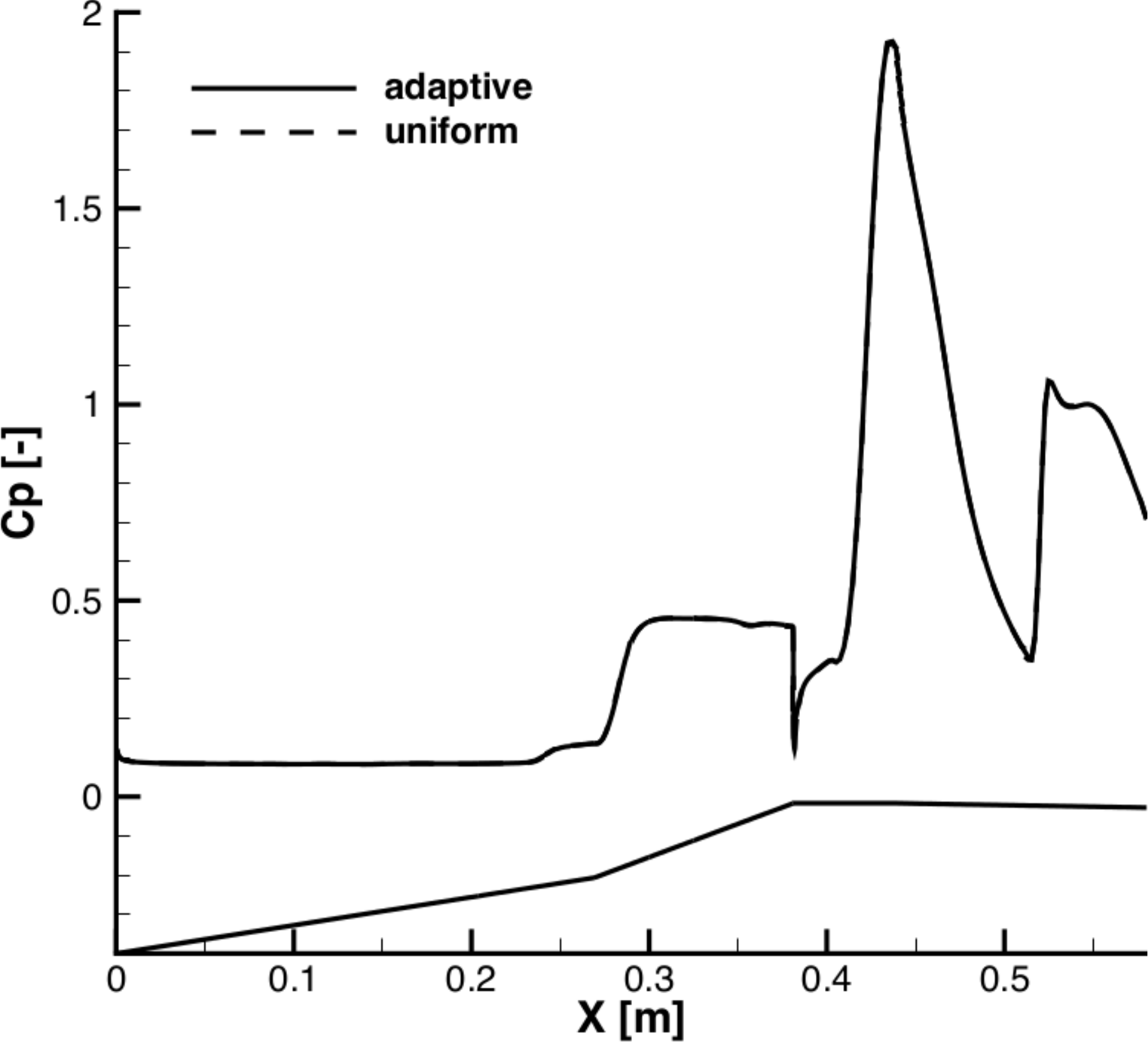}}
    \mbox{\includegraphics [width = 0.5\textwidth,clip]{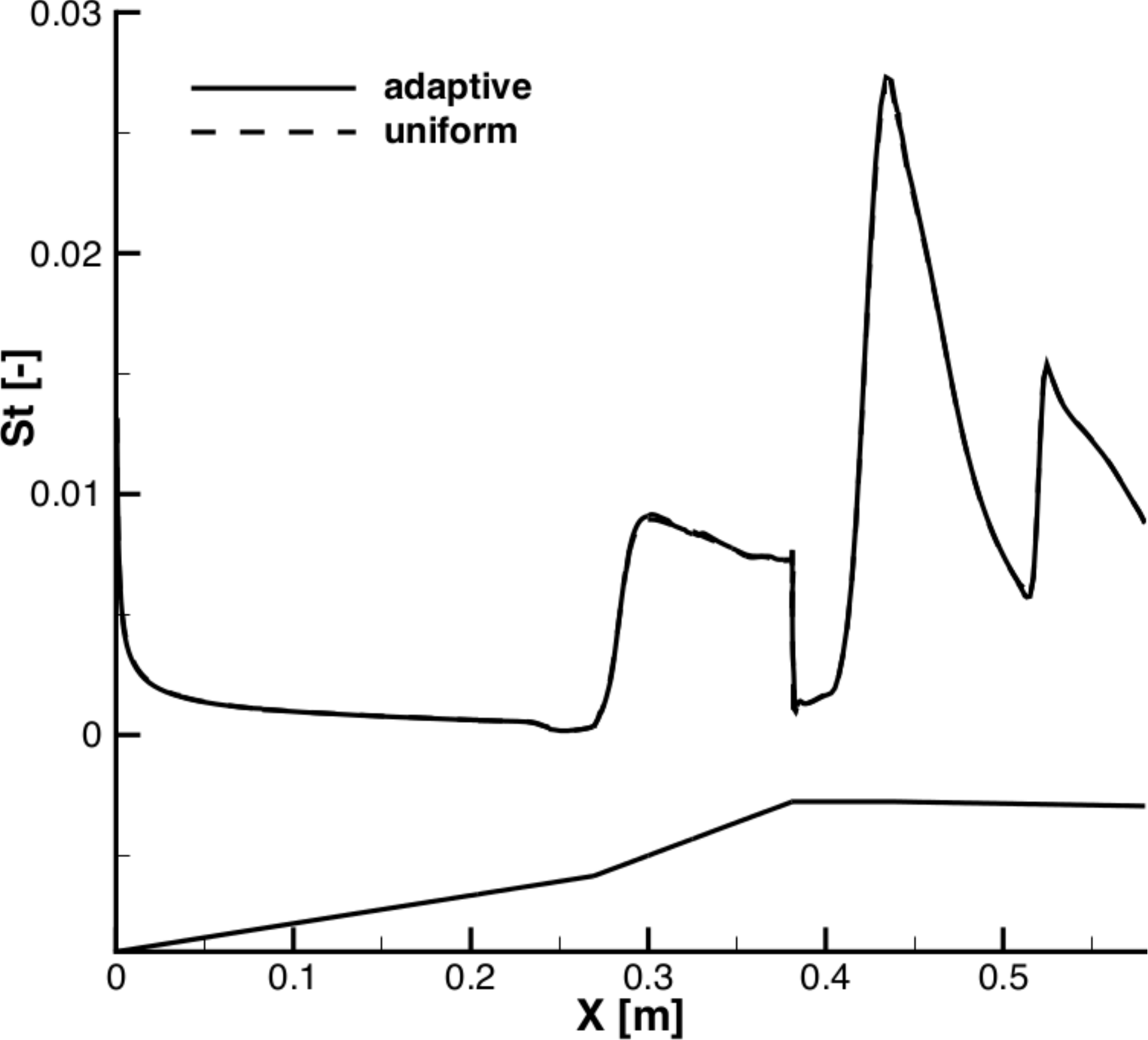}} }
        \caption{Pressure coefficient (left) and Stanton number (right) distribution at lower wall of the scramjet intake in the symmetry plane (z=100\%). Comparison between two adaptive grids using 4 and 5 refinement levels and a uniform refined grid at L=4. Inflow conditions: Re= 4.1$\times$10$^{6}$ 1/m, M= 7.7.}\label{3d-results-scram-z100}
  \end{figure*}
   \begin{figure*}[h!]
 \centerline{
  \includegraphics [width = 0.5\textwidth,clip]{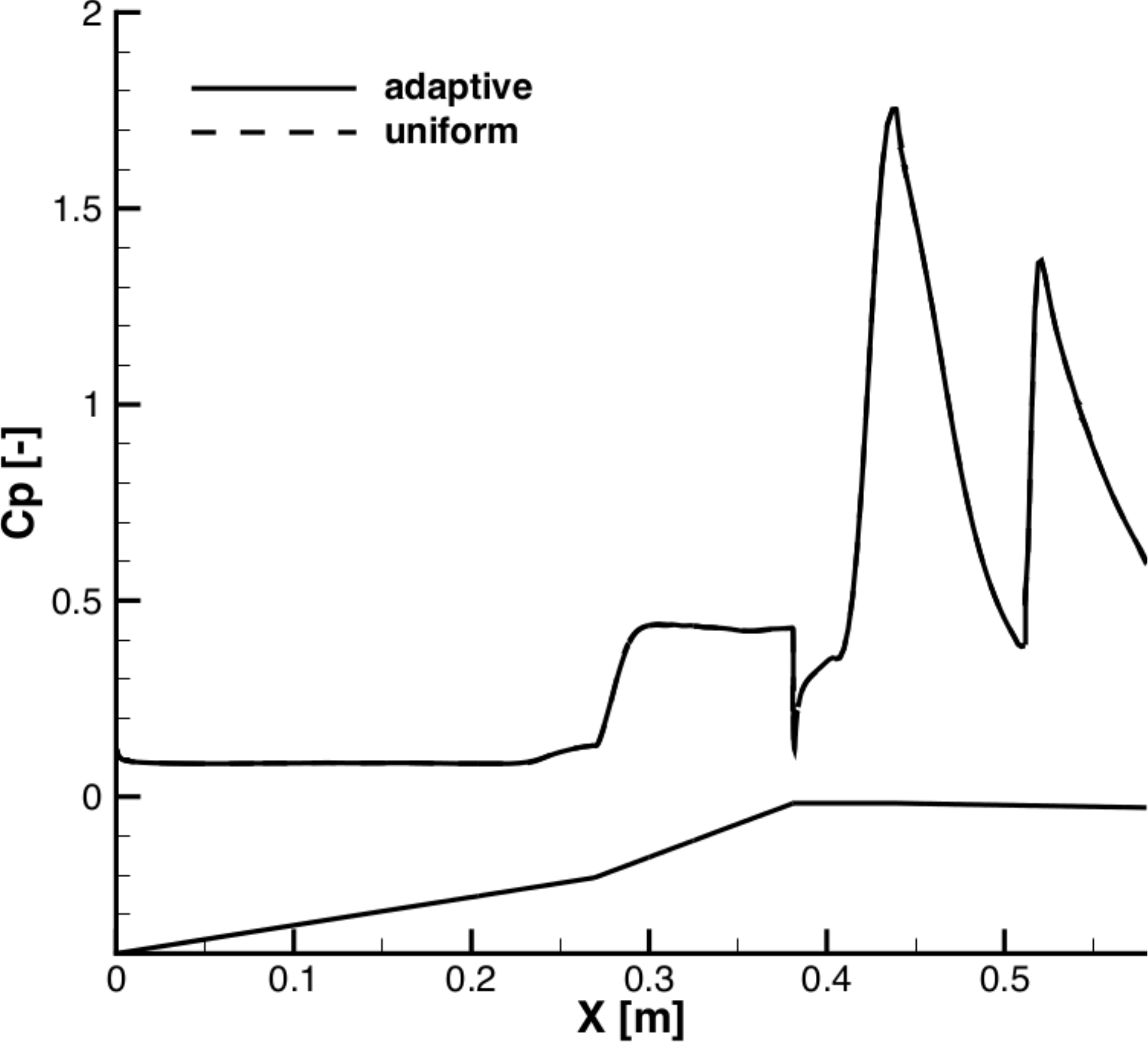}
 \includegraphics [width = 0.5\textwidth,clip]{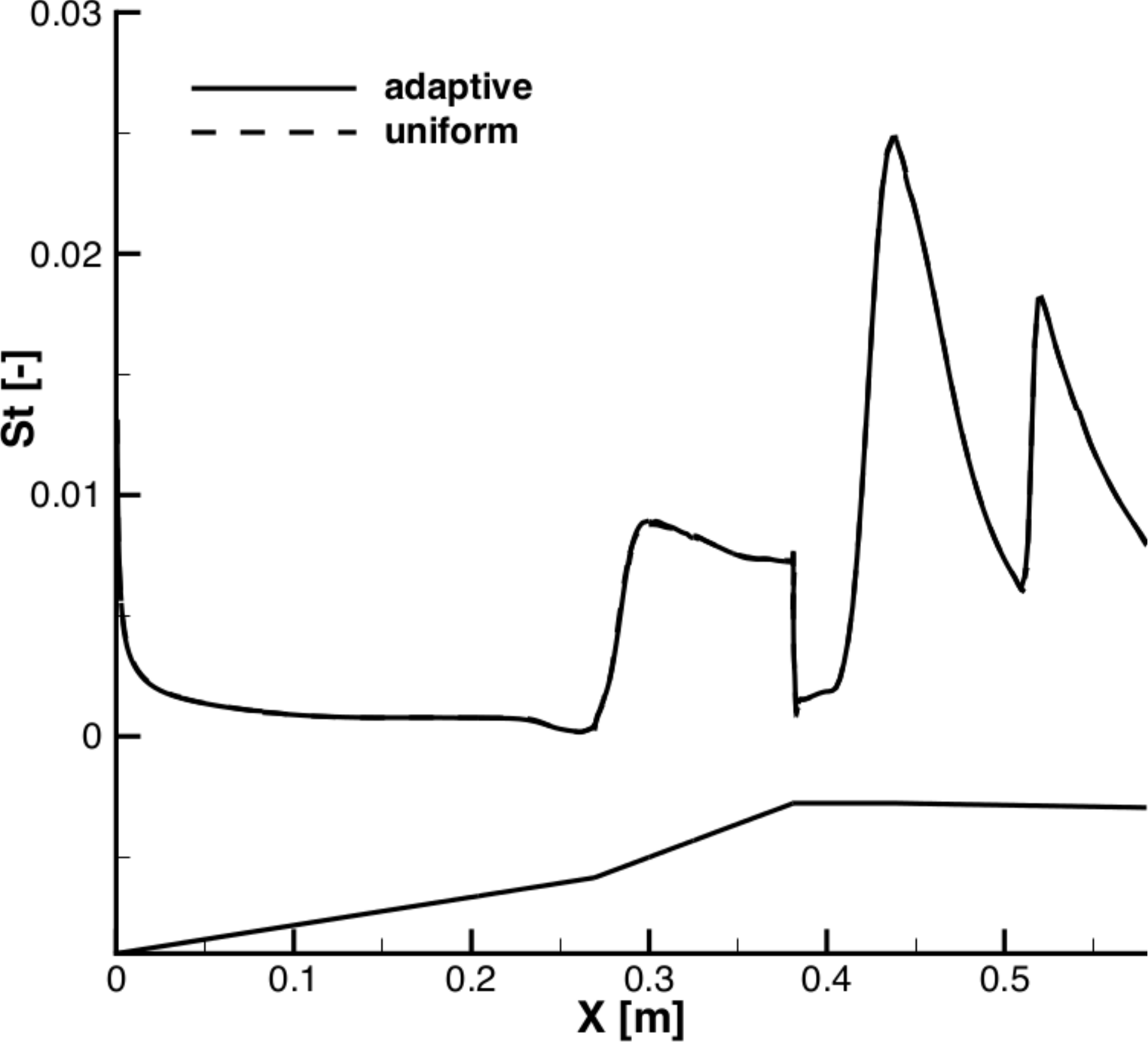} }
        \caption{Pressure coefficient (left) and Stanton number (right) distribution at lower wall of the scramjet intake for z=50\%. Comparison between two adaptive grids using 4 and 5 refinement levels and a uniform refined grid at L=4. Inflow conditions: Re= 4.1$\times$10$^{6}$ 1/m, M= 7.7.}\label{3d-results-scram-z50}
  \end{figure*}
Figures \ref{3d-results-scram-z100} and \ref{3d-results-scram-z50} show the pressure coefficient and Stanton number distributions at the wall along the centerline ($z=100\%$) and for $z=50\%$. This is done to prove that the results obtained by means of grid adaptation are in excellent agreement with the one obtained on a uniform grid for both quantities. 
We note that the results obtained for the adaptive and the uniform grid at level $L=4$ are almost indistinguishable and that no appreciable differences can be seen. 

     \begin{figure*}[h!]
 \centerline{
 \includegraphics [width = 0.5\textwidth,clip]{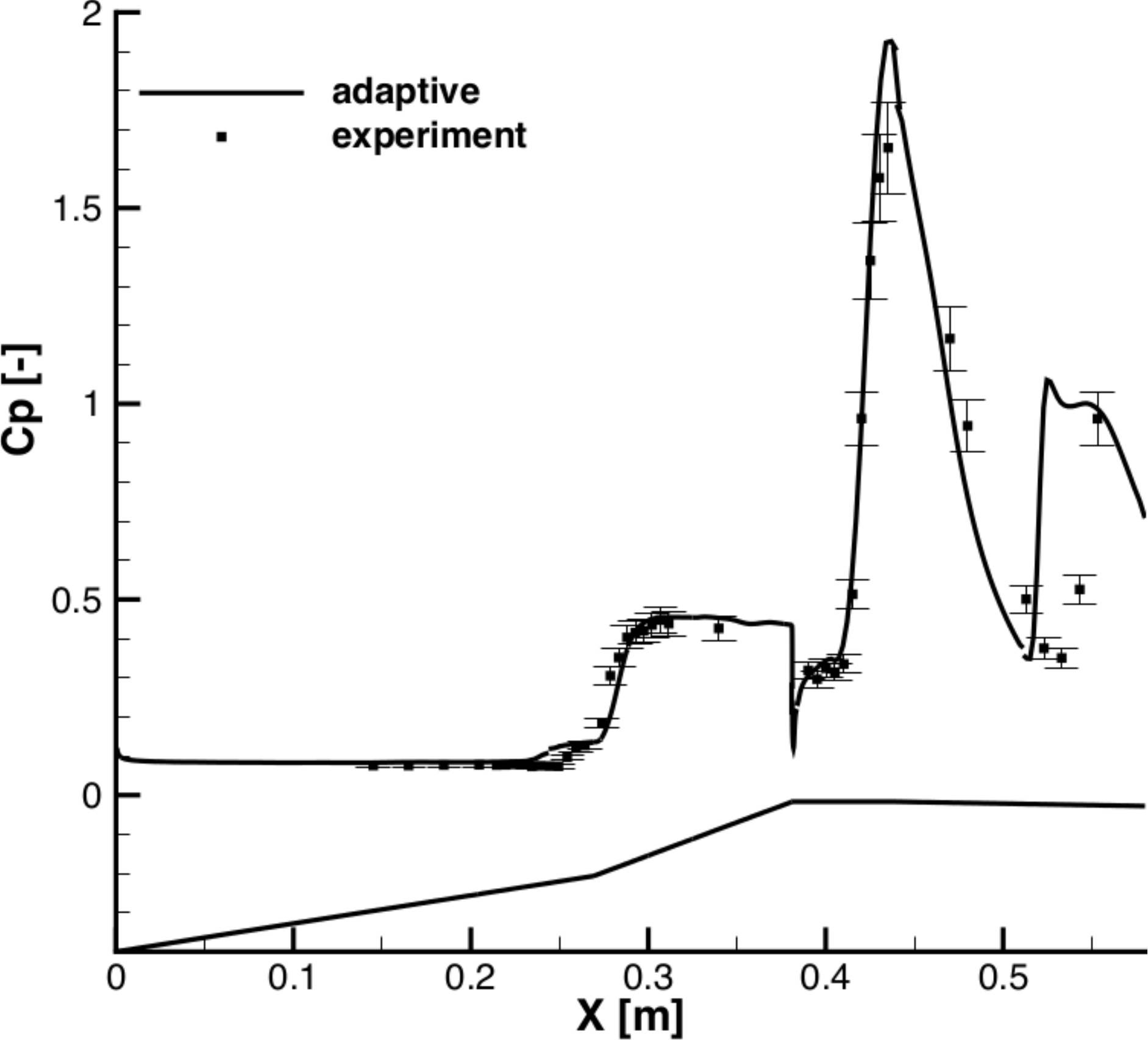}
  \includegraphics [width = 0.5\textwidth,clip]{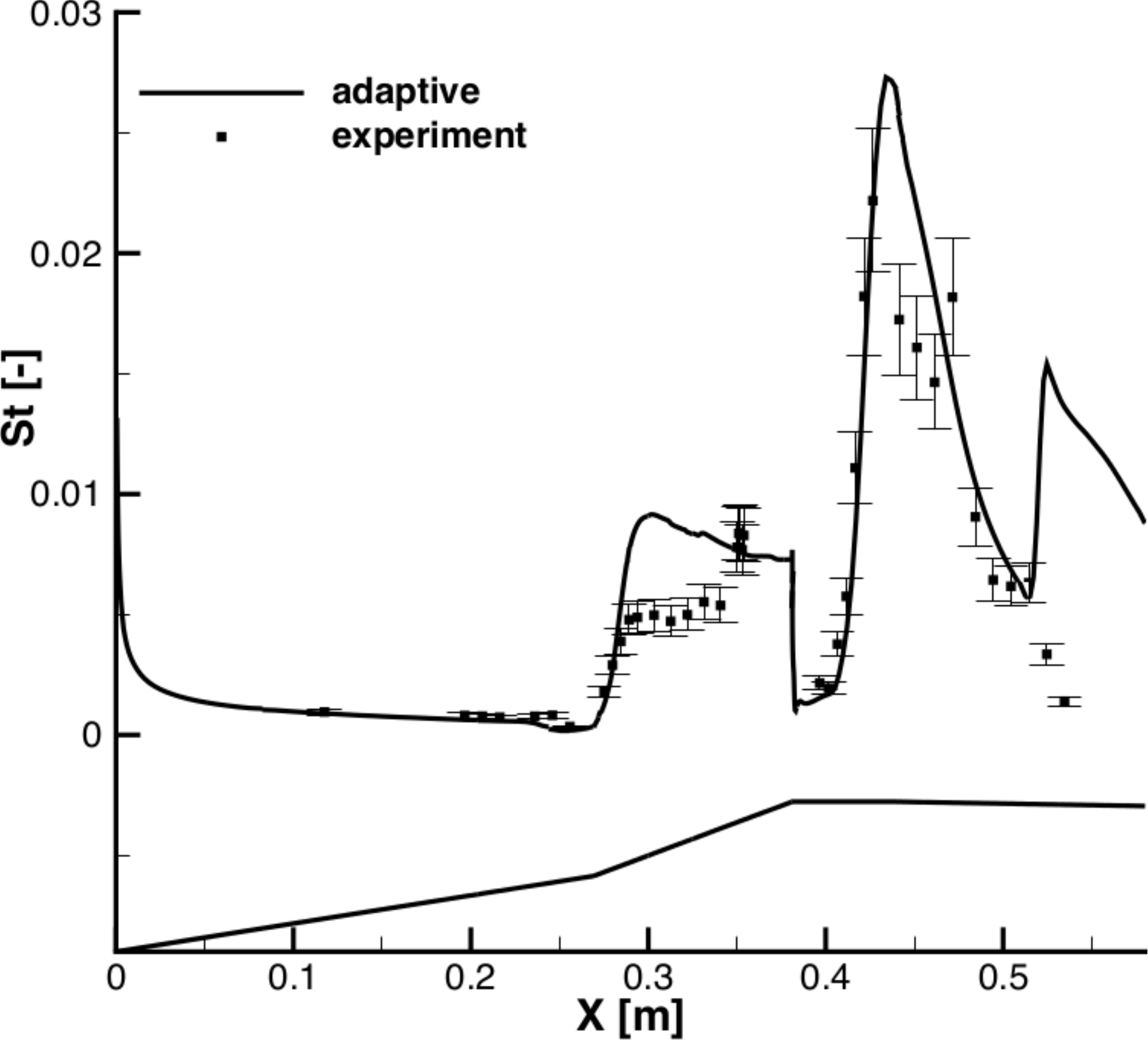}} 
        \caption{Pressure coefficient (left) and Stanton number (right) distribution at lower wall of the scramjet intake in the symmetry plane (z=100\%). Comparison between the adaptive grid using 4 refinement levels and experimental data. Inflow conditions: Re= 4.1$\times$10$^{6}$ 1/m, M= 7.7.}\label{3d-wall-scram}
  \end{figure*}
  
A comparison of the adaptive level $L=4$ results to experimental data \cite{Neuen:06, Fischer} is shown in Figure \ref{3d-wall-scram} for the pressure coefficient and the Stanton number of the lower intake wall along the center line. On the first ramp, the flow is still laminar and the numerical solution follows the experimental data closely.  Within the shear layer over the separation between the two compression ramps the flow becomes transitional. The  ``laminar box'' method (i.e., turning off the turbulence model on the first ramp) used in the numerical approach slightly overpredicts the separation and the pressure coefficient starts to increase at $x=0.23$ m. The measured Stanton numbers after reattachment indicate a transitional behaviour of the flow that is not simulated by the ``laminar box'' method. However, further downstream along the second ramp, the flow becomes indeed turbulent and numerical and experimental distributions agree once again. During the following expansion, where the flow turns inward into the interior engine section, the flow partially relaminarizes. Previous studies have shown that the RSM turbulence model is generally able to predict the relaminarization \cite{Nguyen:2012}. This is important, because the state of the boundary layer needs to be accurately predicted in order to obtain the correct separation size. The separation is caused by the impingement of the cowl shock wave from the upper engine wall. The experimental values are closely matched by the numerical simulation for the separated flow area and the subsequent reattachment peak, although the measured heat flux values within the peak heating area are difficult to interpret. The second, lower peak in the pressure coefficient and the Stanton number is due to the impingement of the reflected reattachment shock, where the reflection occurs on the upper engine wall. The numerical solution predicts the impingement too early, likely due to the fact that a fully turbulent boundary layer is assumed on the upper wall right from the leading edge. In the experiment, we would assume the cowl boundary layer to start laminar at the leading edge. The same discrepancies to the experiments were found for non-adaptive computations on structured grids \cite{Nguyen:10}. Hence the differences are not caused by the adaptive procedure. 

   \begin{figure*}[h!]
      \centerline{
    \includegraphics [width = 1.0\textwidth,clip]{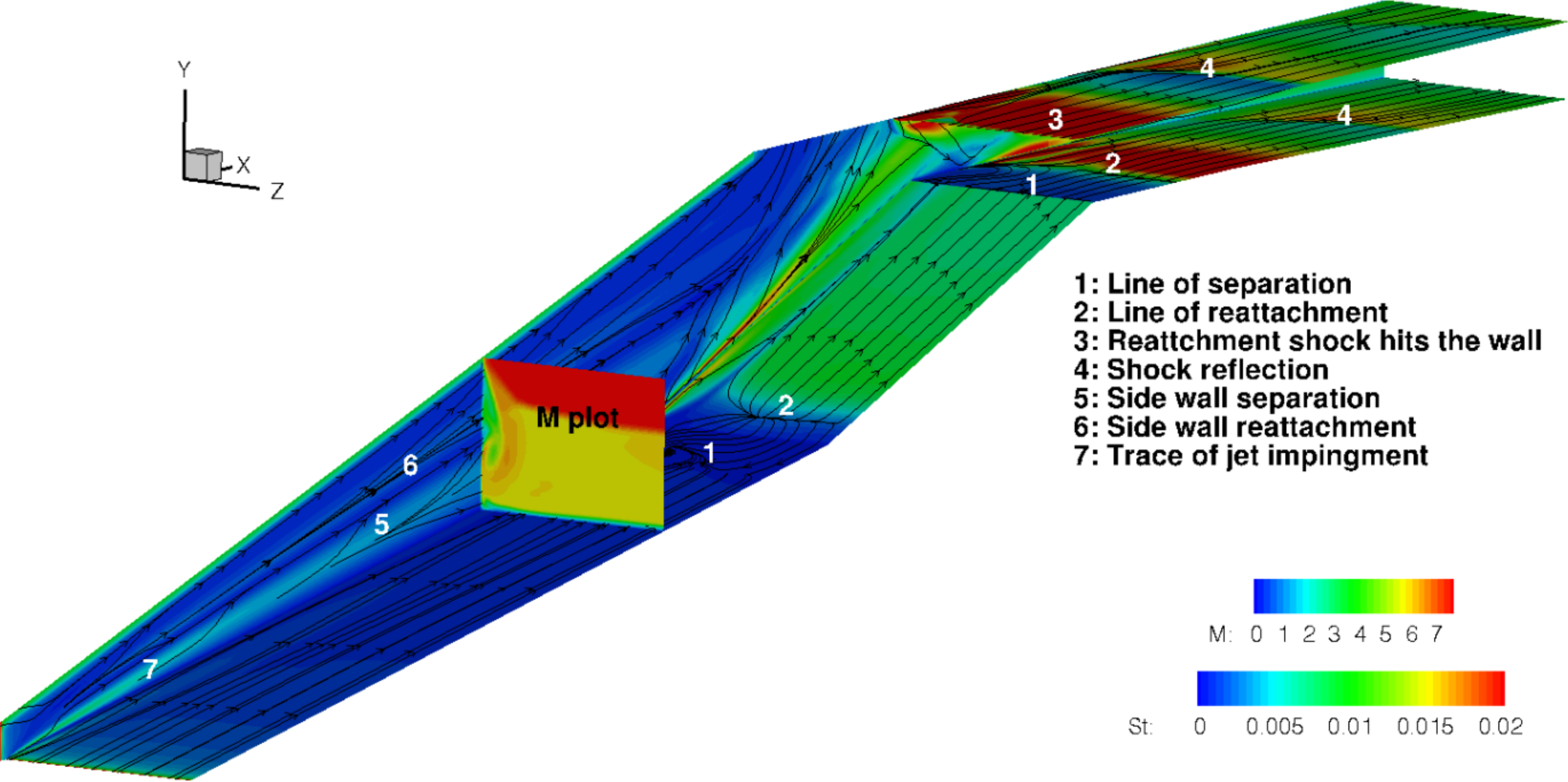} }
        \caption{Heat transfer contours and surface streamlines on the ramps and the side wall. ``M plot'' refers to the Mach number plot as shown in Fig.~\ref{3d-slicex02}. }\label{3d-scram}
  \end{figure*}  
  
To illustrate the three-dimensional effects in the flow field, Figure \ref{3d-scram} shows the footprint of the flow structures on the ramps and the side wall in terms of Stanton number and surface streamlines. 
Also shown is the Mach number distribution in the cross plane at $x=0.2$~m. The separation zone at the kink between the first and second ramp shows the influence of three-dimensional effects because close to the symmetry plane the line of separation (feature (1) in Figure \ref{3d-scram})  extends further upstream than close to the side wall. Also, the reattachment line (2) bends downstream when interacting with the corner flow close to the side wall, making the separation even smaller. At the second ramp and in the expansion region, the flow is mostly two-dimensional. Thus, the surface streamlines are almost parallel. The reattachment shock wave of the first separation bubble hits the upper intake wall and interacts with the cowl shock, causing an area of intense heating at the leading edge of the engine cowl. The boundary layer is here still very thin and no separation is caused. The reflected reattachment shock and the newly generated cowl shock wave form one strong shock wave that impinges on the ramp side. At the ramp, the boundary layer is very thick due to the build--up over two ramps and the subsequent expansion. Hence, a large separation zone is caused by the impinging shock wave, extending all the way to the expansion corner. The separation itself is remarkably two-dimensional  despite the vortices generated by the interaction of the leading edge shock wave and the side wall.
The reattachment shock wave of this second separation (feature (2)  in Figure \ref{3d-scram}) causes subsequent peak heating at the wall due to the compression of the streamlines in this area. The strong reattachment shock impinges on the upper wall and is reflected there, impinging again on the ramp side. Both, upper and lower, impingement points are characterized by a strong increase in the Stanton number. The shock reflections feature a strong three-dimensional effect due to the interaction with the corner flow.

\begin{figure*}[h!]
 \centerline{
 \includegraphics [width = 0.33\textwidth,clip]{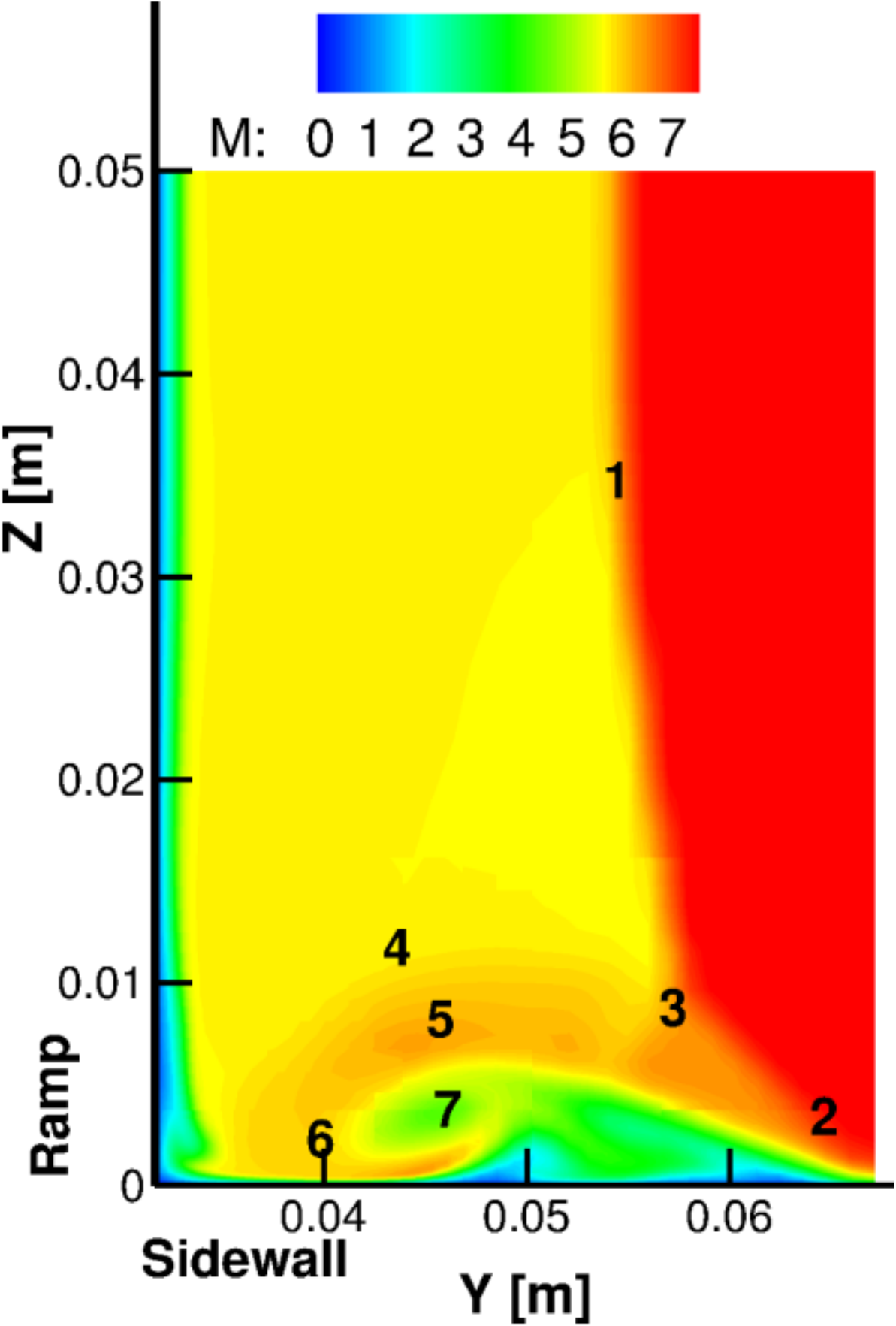}
  \includegraphics [width = 0.33\textwidth,clip]{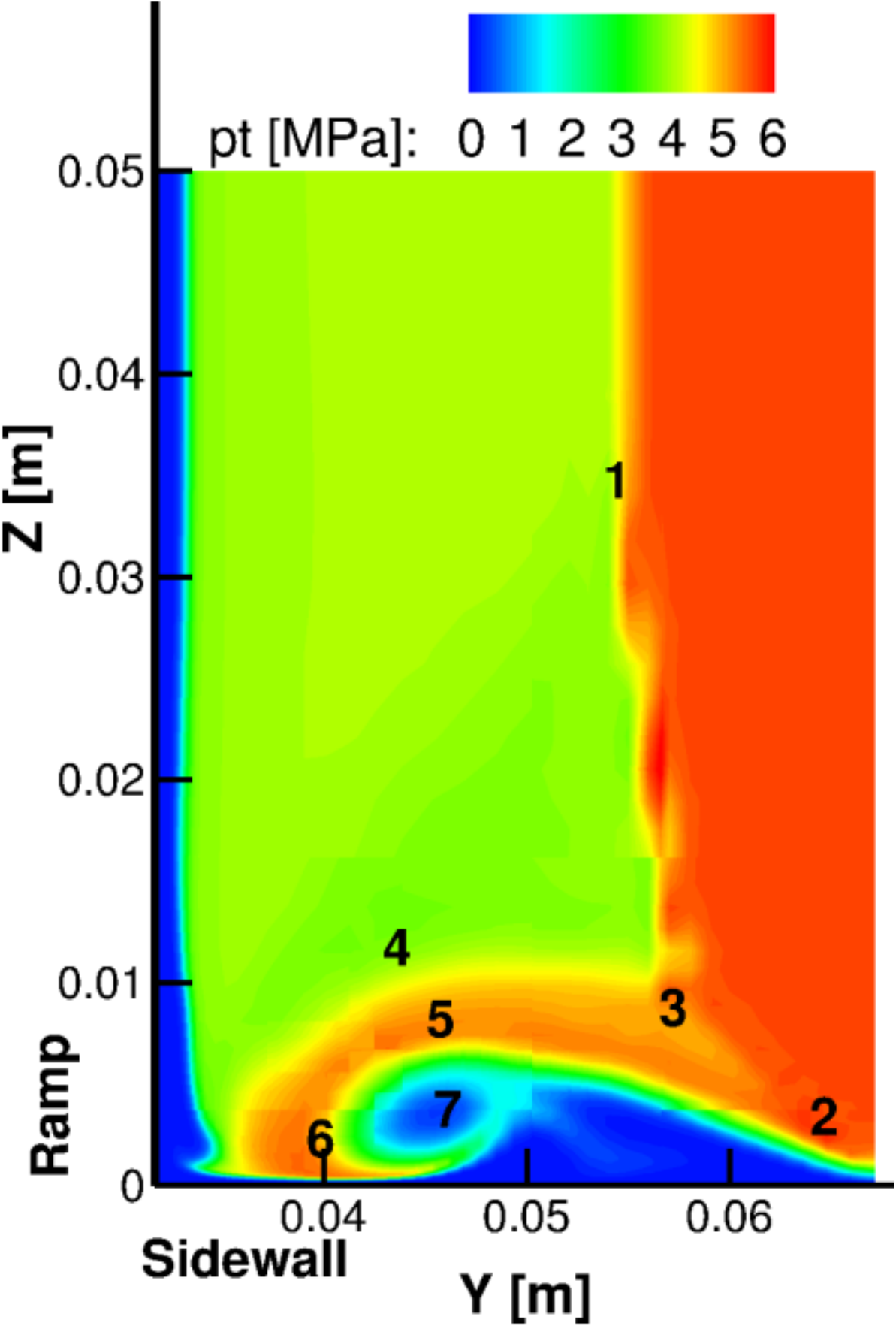}
   \includegraphics [width = 0.33\textwidth,clip]{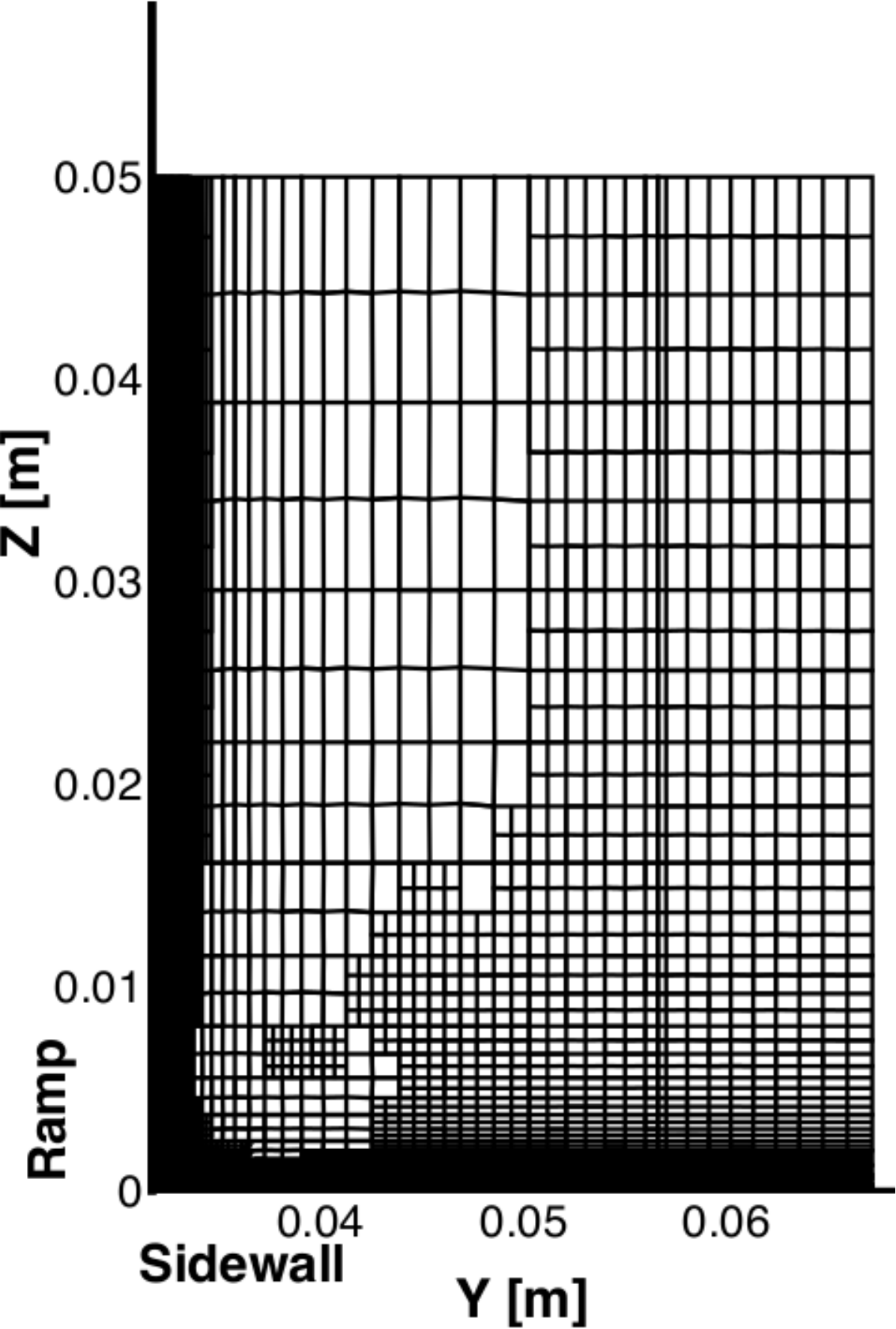}
    }
            \caption{Slice at $x=$0.2~m. Illustration of the flow structures near the sidewall: 1) ramp leading edge shock, 2) separation shock, 3) rear shock, 4) slip line, 5) expansion region, 6) impinging jet, 7) primary vortex. Left: Mach number contour. Middle: Total pressure contour. Right: Final adaptive grid.} \label{3d-slicex02}
  \end{figure*}

The corner flow near the side wall is highly complex due to swept boundary layer interactions occurring in this region. Thus, the correct prediction of the effects is challenging. Close to the side wall, the strong ramp leading edge shock wave interacts with the weak side wall shock and impinges on the boundary layer produced by the side wall. This flow interaction is comparable to the flow around a fin-type configuration for which Alvi and Settles proposed a detailed flow field model \cite{Alvi:91}. All features of this model can be observed in the computation as illustrated in Figure \ref{3d-slicex02}. 
Figure \ref{3d-slicex02} shows the Mach number, the total pressure and the grid for a $zy$-slice at $x=$0.2~m. In the Mach number plot the $\lambda$-shock structure consisting of the separation shock, the rear shock and the ramp leading edge shock as well as the primary vortex and the expansion region are clearly visible. The slip line and the impinging jet are more obvious in the total pressure plot. The grid plot shows that the adaptive procedure detects and resolves all theses flow phenomena correctly. The corner interaction leads to a separation of the flow on the side wall, depicted in Figure \ref{3d-scram}
by the line of separation (5) and reattachment (6). The footprint of the generated jet (feature (6) in Figure \ref{3d-slicex02} ) is also visible by the reduced wall heat flux (7). Remarkable is how well the adaption technique is resolving all of these very different flow phenomena. 

\section{Conclusions}

In this paper, the possibility of studying hypersonic turbulent flows using a differential Reynolds stress turbulence model with a multiscale grid adaptation has been considered. The chosen approach allows to start the simulations on a truly coarse grid that is \emph{not} resolving the physical features of the flow.
By means of the adaptive procedure, the grid is progressively refined and all the important flow features, e.g., shock waves, boundary layers, shear layers, and vortices, are
\emph{automatically} detected and properly resolved. The adaptive technique has proven to be numerically robust also in combination with second order closure turbulence models which are characterized by a decreased numerical stability. 

From a point of view of the performance, the reduction of the grid size leads to a decrease of the number of iterations and to a decrease of the time necessary to perform each iteration. Quantitatively, the CPU time for the three-dimensional simulation of a scramjet intake can be reduced to one third with respect to a uniform grid on the highest refinement level. For two-dimensional simulations, the CPU time is reduced even further, to one eighth or better. This gain is obtained with no loss in the accuracy of the solution in terms of pressure and Stanton number at the wall.
The multiscale grid adaptation can be combined with any standard finite volume solver as described in \cite{Bramkamp:04}.
Depending on the underlying discretization, appropiate wavelets have to be constructed. A general construction procedure is described in \cite{Mueller:03} for arbitrary cell topologies. 

\section*{Acknowledgments}
This work was supported by the German Research Foundation (DFG) within the framework of the GRK 1095 ``Aero-Thermodynamic Design of a Scramjet Propulsion System for Future Space Transportation Systems'' and the GSC 111 Aachen Institute for Advanced Study in Computational Engineering Science (AICES). Computing resources were provided by the RWTH Aachen University Center for Computing and Communication and the Forschungszentrum J\"ulich.

\section*{References}
\bibliography{flower}

\begin{thebibliography}{10}
\newcommand{\enquote}[1]{``#1''}

\bibitem{martin}
Smits, A., Martin, P., and Girimaji, S., \enquote{Current Status of Basic
  Research in Hypersonic Turbulence}, AIAA Paper 2009--0151, January 2009.

\bibitem{Cand:00}
Candler, G., Nompelis, I., and Holden, M., \enquote{Computational Analysis of
  Hypersonic Laminar Vicous-Inviscid Interactions}, {\em AIAA Paper\/}, , No.
  2000--0532, 2000.

\bibitem{Rein:07b}
Reinartz, B., Ballmann, J., Brown, L., Fischer, C., and Boyce, R.,
  \enquote{Shock Wave / Boundary Layer Interaction in Hypersonic Intake Flows},
  {\em 2nd European Conference on Aero-Space Sciences (EUCASS), Brussels,
  Belgium 1-6 July 2007\/}, 2007.

\bibitem{Mack:02}
Mack, A. and Hannemann, V., \enquote{Validation of the Unstructured
  DLR-TAU-Code for Hypersonic Flows}, AIAA Paper 2002-3111, 2002, 32nd AIAA
  Fluid Dynamics Conference and Exhibit, 24-26 June 2002, St.~Louis, Missouri,
  USA.

\bibitem{tau2}
Kovar, A., Hannemann, V., Karl, S., and Sch\"ulein, S., \enquote{About the
  Assessment of Heat Flux and Skin Friction of the DLR TAU-code for Turbulent
  Supersonic Flows}, 8$^{th}$ onera-dlr aerospace symposium conference
  proceedings, 2007.

\bibitem{GottschlichMueller-Mueller:99}
{Gottschlich--M{\"u}ller}, B. and M{\"u}ller, S., \enquote{Adaptive Finite
  Volume Schemes for Conservation Laws based on Local Multiresolution
  Techniques}, {\em Hyperbolic Problems: Theory, Numerics, Applications\/},
  edited by M.~Fey and R.~Jeltsch, Birkh{\"a}user, 1999, pp. 385--394.

\bibitem{harten}
Harten, A., \enquote{Multiresolution algorithms for the numerical solution of
  hyperbolic conservation laws}, {\em Comm.\ Pure Appl.\ Math.\/}, Vol.~48,
  No.~12, 1995, pp.~1305--1342.

\bibitem{mueller:09}
M\"uller, S., \enquote{Multiresolution Schemes for Conservation Laws}, {\em
  Multiscale, Nonlinear and Adaptive Approximation\/}, edited by R.~DeVore and
  A.~Kunoth, Springer-Verlag, 2005, pp. 379--408.

\bibitem{Bramkamp:04}
Bramkamp, F.~D., Lamby, P., and M\"uller, S., \enquote{An adaptive multiscale
  finite volume solver for unsteady and steady flow computations}, {\em Journal
  of Computational Physics\/}, Vol.~197, 2004, pp.~460--490.

\bibitem{Bramkamp:03b}
Bramkamp, F., {\em Unstructured h-Adaptive Finite-Volume Schemes for
  Compressible Viscous Fluid Flow\/}, Ph.D. thesis, RWTH Aachen University,
  2003.

\bibitem{SM-Lamby:07}
Lamby, P., {\em Parametric Multi-Block Grid Generation and Application To
  Adaptive Flow Simulations\/}, Ph.D. thesis, RWTH Aachen, 2007.

\bibitem{Mueller:03}
M\"uller, S., \enquote{Adaptive Multiscale Schemes for Conservation Laws}, {\em
  Lecture Notes on Computational Science and Engineering\/}, Vol.~27,
  Springer-Verlag, 2003.

\bibitem{Brix:09}
Brix, K., Mogosan, S., M\"uller, S., and Schieffer, G.,
  \enquote{Parallelization of multiscale-based grid adaptation using
  space-filling curves}, {\em ESAIM Proceedings\/}, edited by F.~Coquel,
  Y.~Maday, S.M\"uller, M.~Postel, and Q.~Tran, Vol.~29, 2009, pp. 108--129.

\bibitem{Brix:11}
Brix, K., Melian, S.~S., M\"uller, S., and Bachmann, M., \enquote{Adaptive
  Multiresolution Methods: Practical Issues on Data Strucutres, Implementation
  and Parallelization}, {\em ESAIM Proceedings\/}, edited by V.~Louvet and
  M.~Massot, Vol.~34, 2011, pp. 151--183.

\bibitem{roy-blottner}
Roy, J.~C. and Blottner, F.~G., \enquote{Review and assessment of turbulence
  models for hypersonic flows}, {\em Progress in Aeropsace Sciences\/},
  Vol.~42, 2006, pp.~469--530.

\bibitem{Bosco:11b}
Bosco, A., {\em Reynolds Stress Model for Hypersonic Flows\/}, Ph.D. thesis,
  RWTH Aachen University, 2011.

\bibitem{Bosco:11}
Bosco, A., Reinartz, B., Brown, L., and Boyce, R., \enquote{Investigation of a
  Compression Corner at Hypersonic Conditions using a Reynolds Stress Model},
  AIAA Paper 2011-2217, April 2011.

\bibitem{Eisfeld:06}
Eisfel, B., \enquote{Computation of Complex Compressible Aerodynamic Flows with
  A Reynolds Stress Turbulence Model}, Int. Conference on Boundary and Interior
  Layers BAIL, G. Lube, G. Rapin (Eds), 24-28 July 2006, G\"ottingen, Germany,
  2006.

\bibitem{SSG}
Speziale, C., Sarkar, S., and Gatski, T., \enquote{Modelling the
  pressure-strain correlation of turbulence: an invariant dynamical system
  approach}, {\em J. of Fluid Mechanics\/}, Vol.~227, 1991, pp.~254--272.

\bibitem{LRR}
Launder, B.~R., Reece, G.~J., and Rodi, W., \enquote{Progress in the
  development of a Reynolds-stress turbulence closure}, {\em J. of Fluid
  Mechanics\/}, Vol.~68, 1975, pp.~537--566.

\bibitem{wi_1}
Wilcox, D.~C., \enquote{Turbulence Energy Equation Models}, {\em Turbulence
  Modeling for CFD\/}, Vol.~1, DCW Industries, Inc., La Canada, CA, 2nd ed.,
  1994, pp. 73--170.

\bibitem{me_1}
Menter, F.~R., \enquote{Two-Equation Eddy-Viscosity Turbulence Models for
  Engineering Applications}, {\em AIAA Journal\/}, Vol.~32, No.~8, Aug. 1994,
  pp.~1598--1605.

\bibitem{Ballmann:03}
Ballmann, J., \enquote{Flow Modulation and Fluid-Structure Interaction at
  Airplane Wings}, {\em Notes on Numerical Fluid Mechanics and
  Multidisciplinary Design 84\/}, Springer Verlag, 2003.

\bibitem{schroeder}
Schr\"oder, W., \enquote{Flow Modulation and Fluid-Structure Interaction
  Findings}, {\em Numerical Fluid Mechanics and Multidisciplinary Design (NNFM)
  109, Springer Verlag\/}, 2010.

\bibitem{limiter}
Venkatakrishnan, V., \enquote{Convergence to {S}teady {S}tate {S}olutions of
  the {E}uler {E}quations on {U}nstructured {G}rids with {L}imiters}, {\em
  Journal of Computational Physics\/}, Vol.~118, 1995, pp.~120--130.

\bibitem{Bramkamp:03}
Bramkamp, F., Gottschlich-M\"uller, B., Hesse, M., Lamby, P., M\"uller, S.,
  Ballmann, J., Brakhage, K., and Dahmen, W., \enquote{H-Adaptive Multiscale
  Schemes for Compressible Navier-Stokes Equations - Polyhedral Discretization,
  Data Compression and Mesh Generation}, {\em Notes on Numerical Fluid
  Mechanics\/}, Vol.~84, 2003, pp.~125--204.

\bibitem{harten96}
Harten, A., \enquote{Multiresolution representation of data: {A} general
  framework}, {\em SIAM J.\ Numer.\ Anal.\/}, Vol.~33, No.~3, 1996,
  pp.~1205--1256.

\bibitem{carnicer}
Carnicer, J., Dahmen, W., and Pe{\~{n}}a, J., \enquote{Local decomposition of
  refinable spaces and wavelets}, {\em Appl.\ Comput.\ Harmon.\ Anal.\/},
  Vol.~3, 1996, pp.~127--153.

\bibitem{Mueller:2003a}
Cohen, A., Kaber, S.~M., M\"uller, S., and Postel, M., \enquote{Fully Adaptive
  Multiresolution Schemes for Conservation Laws}, {\em Mathematics of
  Computation\/}, Vol.~72, 2003, pp.~183--225.

\bibitem{Mueller:2010}
Hovhannisyan, N. and M\"uller, S., \enquote{On the stability of fully adaptive
  multiscale schemes for conservation laws using approximate flux and source
  reconstruction strategies}, {\em IMA Journal of Numerical Analysis\/},
  Vol.~30, 2010, pp.~1256--1295.

\bibitem{Cohen92}
Cohen, A., Daubechies, I., and Feauveau, J., \enquote{Bi-orthogonal bases of
  compactly supported wavelets}, Vol.~45, 1992, pp.~485--560.

\bibitem{haar:10}
Haar, A., \enquote{Zur {T}heorie der orthogonalen {F}unktionen-{S}ysteme}, {\em
  Math.\ Ann.\/}, Vol.~69, 1910, pp.~331 --371.

\bibitem{Frauholz:12}
Frauholz, S., Behr, M., and Reinartz, B., \enquote{Numerical Simulation of
  Hypersonic Air Intake Flow in Scramjet Propulsion Using a Mesh-Adaptive
  Approach}, {\em 18th AIAA International Space Planes and Hypersonic Systems
  and Technologies Conference, Tours, France, 23-27 September 2012, AIAA Paper
  2012-5976\/}, 2012.

\bibitem{Neuen:06}
Neuenhahn, T. and Olivier, H., \enquote{Influence of the wall temperature and
  the entropy layer effects on double wedge shock boundary layer interactions},
  AIAA Paper 2006-8136, 2006.

\bibitem{Fischer}
Fischer, C. and Olivier, H., \enquote{Experimental investigation of the
  internal flow field of a scramjet engine}, AIAA Paper 2009-7369, September
  2008.

\bibitem{Nguyen:10}
Nguyen, T., Schieffer, G., Fischer, C., Olivier, H., Behr, M., and Reinartz,
  B., \enquote{Details of Turbulence Modeling in Numerical Simulations of
  Scramjet Intake}, {\em 27th Congress of International Council of the
  Aeronautical Sciences (ICAS), Nice, France, 19-24 September 2010\/}, 2010.

\bibitem{Gaisbauer:07}
Gaisbauer, U., Weigand, B., and Reinartz, B., \enquote{Research Training Group
  GRK 1095/1: Aero-Thermodynamic Design of a Scramjet Propulsion System}, {\em
  Proceedings of 18th ISABE Conference\/}, International Symposium of
  Air-Breathing Engines (ISABE), Beijing, China, September 2-7, 2007.

\bibitem{Nguyen:2012}
Nguyen, T., Vukovic, M., Behr, M., and Reinartz, B., \enquote{Numerical
  Simulation of Successive Distortions in Supersonic Turbulent Flows}, {\em
  AIAA Journal\/}, Vol.~50, No.~11, November 2012, pp.~2365--2375.

\bibitem{Alvi:91}
Alvi, F.~S. and Settles, G.~S., \enquote{A Physical Model of the Swept
  Shock/Boundary-Layer Interaction Flowfield}, {\em AIAA 22nd Fluid Dynamics,
  Plasma Dynamics and Laser Conference\/}, Honolulu, Hawaii, June 1991.

\end{thebibliography}
\bibliographystyle{aiaa}

\end{document}